\documentstyle[epsfig,natbib2,amsmath,natbibmnfix]{mn}

\newcommand{\be}{\begin{equation}}
\newcommand{\ee}{\end{equation}}
\newcommand{\bea}{\begin{eqnarray}}
\newcommand{\eea}{\end{eqnarray}}
\newcommand{\bc}{\begin{center}}
\newcommand{\ec}{\end{center}}

\renewcommand{\vec}[1]{ {\bmath #1} }

\newcommand{\Gyr}{\,\unit{Gyr}}

\renewcommand{\thefootnote}{\fnsymbol{footnote}}

\newcommand{\unit}[1]{\text{#1}}
\newcommand{\diff}[1]{\unit{d}#1}

\newcommand{\nablavec}{\vec{\nabla}}

\title{Thermal conduction in cosmological SPH simulations}

\author[M.~Jubelgas, V.~Springel, K.~Dolag] 
{\parbox{18cm}
  {
    Martin Jubelgas$^{1}$\footnotemark[1],
    Volker Springel$^{1}$ and
    Klaus Dolag$^2$
  }
  \vspace{0.3cm}\\
  $^1$Max-Planck-Institut
  f\"{u}r Astrophysik, Karl-Schwarzschild-Stra\ss{}e 1, 85740 Garching
  bei M\"{u}nchen, Germany\\ 
  $^2$Dipartimento di Astronomia, Universit\`a di Padova, Viccolo
  dell'~Osservatorio 5, 1-35122 Padova, Italy
}

\begin{document}

\maketitle

\begin{abstract}
  Thermal conduction in the intracluster medium has been proposed as a
  possible heating mechanism for offsetting central cooling losses in
  rich clusters of galaxies. However, because of the coupled
  non-linear dynamics of gas subject to radiative cooling and thermal
  conduction, cosmological hydrodynamical simulations are required to
  reliably predict the effects of heat conduction on structure
  formation.  In this study, we introduce a new formalism to
  model conduction in a diffuse ionised plasma using smoothed particle
  hydrodynamics (SPH), and we implement it in the parallel
  TreePM/SPH-code {\small GADGET-2}. We consider only isotropic
  conduction and assume that magnetic suppression can be described in
  terms of an effective conductivity, taken as a fixed fraction of the
  temperature-dependent Spitzer rate. We also account for saturation
  effects in low-density gas.  Our formulation manifestly conserves
  thermal energy even for individual and adaptive timesteps, and is
  stable in the presence of small-scale temperature noise. This allows
  us to evolve the thermal diffusion equation with an explicit time
  integration scheme along with the ordinary hydrodynamics.  We use a
  series of simple test problems to demonstrate the robustness and
  accuracy of our method.  We then apply our code to spherically
  symmetric realizations of clusters, constructed under the
  assumptions of hydrostatic equilibrium and a local balance between
  conduction and radiative cooling. While we confirm that conduction
  can efficiently suppress cooling flows for an extended period of
  time in these isolated systems, we do not find a similarly strong
  effect in a first set of clusters formed in self-consistent
  cosmological simulations.  However, their temperature profiles are
  significantly altered by conduction, as is the X-ray luminosity. 
\end{abstract}

\begin{keywords}
  galaxies: clusters: general -- conduction -- methods: numerical.
\end{keywords}

\section{Introduction}

\renewcommand{\thefootnote}{\fnsymbol{footnote}}
\footnotetext[1]{E-mail: jubelgas@mpa-garching.mpg.de}
  
Clusters of galaxies provide a unique laboratory to study structure
formation and the material content of the Universe, because they are not
only the largest virialized systems but are also believed to
contain a fair mixture of cosmic matter. Among the many interesting aspects
of cluster physics, their X-ray emission takes a particularly
prominent role.  It provides direct information on the thermodynamic state
of the diffuse intracluster gas, which makes up for most of the baryons in
clusters.

While the bulk properties of this gas are well understood in terms of
the canonical $\Lambda$CDM model for structure formation, there are a
number of discrepancies between observations and the results of
present hydrodynamical simulations. For example, a long standing
problem is to understand in detail the scaling relations of observed
clusters, which deviate significantly from simple self-similar
predictions.  In particular, poor clusters of galaxies seem to contain
gas of higher entropy in their centres than expected
\citep{Ponman99,LD00}.  This has been interpreted either to be
evidence for an entropy injection due to non-gravitational processes
\citep{Loewen00,wu99,Metzler94}, or as a sign of the selective removal
of low-entropy gas by gas cooling \citep{Voit02,Wu02}. Both processes
combined could influence the thermodynamic properties of the ICM in a
complex interplay \citep{Tornatore03,Borgani03}.

Another interesting problem occurs for the radial temperature profiles
of clusters.  Most observed clusters show a nearly isothermal
temperature profile, often with a smooth decline in their central
parts \citep{Allen01,Johnstone02,Ettori02}. Nearly isothermal profiles
are also obtained in adiabatic simulations of cluster formation
\citep[e.g.][]{Fr98}. However, clusters in simulations that include
dissipation typically show temperature profiles that increase towards
the centre \citep[e.g.][]{Lew00}, quite different from what is
observed.

Perhaps the biggest puzzle is that spectroscopic X-ray
observations of the centres of clusters of galaxies have revealed
little evidence for cooling of substantial amounts of gas out of the
intracluster medium \citep[e.g.][]{David2001}, even though this would be
expected based on their bolometric X-ray luminosity alone
\citep{Fabian1994}.  The apparent absence of strong cooling flows in
clusters hence indicates the presence of some heating source for the
central intracluster medium.  Among the proposed sources are AGN,
buoyant radio bubbles \citep{Churazov2001,Ensslin2002}, feedback
processes from star formation \citep{Bower01,Menci00}, or acoustic
waves \citep{Fujita2003}.

Recently, \citet{Narayan01} have proposed that thermal conduction may
play an important role for the cooling processes in clusters.  The
highly ionised hot plasma making up the ICM in rich clusters of
galaxies should be efficient in transporting thermal energy,
unless heat diffusion is inhibited by magnetic fields. If conduction
is efficient, then cooling losses in the central part could be offset
by a conductive heat flow from hotter outer parts of clusters, which
forms the basis of the conduction idea.

Indeed, using simple hydrostatic cluster models where cooling and
conductive heating are assumed to be locally in equilibrium,
\citeauthor{Zak03}~(\citeyear{Zak03}, ZN henceforth) have shown that
the central temperature profiles of a number of clusters can be well
reproduced in models with conduction \citep[see
also][]{Fabian2002,Voigt2002,Brueggen2003,Voigt2003}. The required
conductivities are typically sub-Spitzer \citep{Medvedev2003}.
This suggests that thermal conduction may play an important role for
the thermodynamic properties of the ICM.

On the other hand, it has been frequently argued
\citep{Chandran98,Malyshkin01} that magnetic fields in clusters
most likely suppress the effective conductivity to values well below
the Spitzer value for an unmagnetised gas. Note that
rotation measurements show that magnetic fields {\em do exist} in
clusters \citep[e.g.][]{Vogt2003}. However, little is known about the
small-scale field configuration, so that there is room for models with
chaotically tangled magnetic fields \citep{Narayan01}, which may leave
a substantial fraction of the Spitzer conductivity intact.  The
survival of sharp temperature gradients along cold fronts, as observed
by Chandra in several clusters
\citep{Markevitch00,Vikhlinin01,Ettori00}, may require an
ordered magnetic field to suppress
conduction. For a more complete review of the effects of
magnetic fields on galactic clusters, see \cite{Carilli02} and
references therein.

It is clearly of substantial interest to understand
in detail the effects conduction may have on the formation and
structure of galaxy clusters.  In particular, it is far from clear
whether the temperature profile required for a local balance between
cooling and conduction can naturally arise during hierarchical
formation of clusters in the $\Lambda$CDM cosmology.  This question is
best addressed with cosmological hydrodynamical simulations that fully
account for the coupled non-linear dynamics of the gas subject to
radiative cooling and thermal conduction.

In this paper, we hence develop a new numerical implementation of
conduction and include it in a modern TreeSPH code for structure
formation.  Smoothed particle hydrodynamics \citep{Lu77,Mo85,Mo92} is
a powerful numerical tool to investigate gas dynamics, which has
found widespread application in astrophysics.  In cosmological
simulations, gas densities vary over many orders of magnitude, and can
change rapidly as function of position and time. Classical
mesh-based Euclidean approaches to hydrodynamics have difficulty
adjusting to this high dynamic range unless sophisticated adaptive
mesh refinements (AMR) methods are used.
In contrast, the Lagrangian approach of SPH, thanks to its
matter-tracing nature, guarantees good spatial resolution
in high-density regions, while spending little computational time on
low-density regions of space, where a coarser spatial resolution is
usually sufficient.

Another useful aspect of SPH is that the equations that describe
physical processes can often be translated into an SPH form rather
intuitively. Below we discuss this in detail for the conduction
equation, highlighting in particular how a number of practical
problems with respect to stability can be overcome.  The final
formulation we propose is robust and explicitly conserves
thermal energy, even when individual timesteps for each particle are
used.

After examining idealised test problems to validate our implementation
of conduction, we apply our code to realizations of the static cluster
models of ZN, investigating in particular, to what degree conduction
may balance cooling in these clusters, and for how long this
approximate equilibrium can be maintained.  We also discuss results for
a first set of cosmological simulations of cluster formation.  We here
compare simulations that follow only adiabatic gas dynamics, or only
cooling and star formation, with corresponding ones that also include
conduction. As we will see, thermal conduction can lead to a
substantial modification of the final thermodynamic properties of
rich clusters.

The outline of this paper is as follows.  In Section~2, we discuss the
basics of the conduction equation, and our numerical approach for
discretising it in SPH. In Section~3, we show first test runs of
conducting slabs, which we also use to illustrate various issues of
numerical stability.  In Section~4, we consider isolated clusters,
constructed with an initial equilibrium model, while in Section~5 we
present a comparison of results obtained in cosmological simulations
of cluster formation.  Finally, we summarise and discuss our findings
in Section~6.

\section{Thermal conduction in SPH}

\subsection{The conduction equation}

Heat conduction is a transport process for thermal energy, driven by
temperature gradients in the conducting medium. Provided the mean free path
of particles is small compared to the scale length of the temperature
variation, the local heat flux can be described by
\begin{equation}
\vec{j} = - \kappa \nablavec T ,
\label{eqn:heatflux}
\end{equation}
where $T(\vec{r})$ gives the temperature field, and $\kappa$ is
the heat conduction coefficient, which may depend on local properties
of the medium. For example, in the case of an astrophysical plasma, we
encounter a strong dependence of $\kappa$ on the temperature itself.

The rate of temperature change induced by conduction can simply
be obtained from conservation of energy, viz.
\begin{equation}
\rho \frac{\diff{u}}{\diff{t}} = - \nablavec {\boldsymbol \cdot} \vec{j},
\end{equation}
where $u$ is the thermal energy per unit mass, and $\rho$ denotes gas
density. Eliminating the heat flux, this can also be written directly
in terms of $u$, giving the heat conduction equation in the form
\begin{equation}
\frac{\diff{u}}{\diff{t}} = \frac{1}{\rho} \nablavec {\boldsymbol \cdot} ( \kappa
\nablavec T ) \text{.}
\label{eqn:conduction}
\end{equation}
Note that the temperature is typically simply proportional to $u$, unless
the mean particle weight changes in the relevant temperature regime, for
example as a result of a phase transition from neutral gas to ionised
plasma.  In the thin astrophysical plasmas we are interested in, the
strong temperature dependence of the Spitzer conductivity (see below)
suppresses
conduction in low-temperature gas heavily. For all practical
purposes we can set $u=k T /[(\gamma-1)\mu] = c_v T$, where
$\mu=0.588\,m_{\rm p}$ is the mean molecular weight of a fully ionised gas
with the primordial mix of helium and hydrogen, and $c_v$ is the heat
capacity per unit mass.

\subsection{Spitzer Conductivity}

\cite{Spitzer62} derived the classical result for the heat conductivity due
to electrons in an ionised plasma. It is given by \be \kappa_{\rm sp} =
1.31\, n_e \lambda_e k \left(\frac{k T_e}{m_e}\right)^{1/2}, \ee where
$n_e$ is the electron density, and $\lambda_e$ the electron mean free path.
Interestingly, the product $n_e \lambda_e$ depends only on the electron
temperature $T_e$, \be \lambda_e n_e = \frac{3^{3/2} (k T_e)^2}
{4\sqrt{\pi} e^4 \ln \Lambda}, \ee provided we neglect the very weak
logarithmic dependence of the Coulomb logarithm $\Lambda$ on electron
density and temperature, which is a good approximation for clusters. We
will set $\ln \Lambda=37.8$, appropriate for the plasma in clusters of
galaxies \citep{Sarazin1988}. The Spitzer conductivity then shows only a
strong temperature dependence, $\kappa_{\rm sp}\propto T^{5/2}$, and has
the value
\begin{equation}
  \kappa_{\rm sp} = 8.2 \times 10^{20} \left( \frac{ k_{\rm B} T }{10\,
  \unit{keV}} \right)^{5/2} \frac{\unit{erg}}{\unit{cm}\ \unit{s}\
  \unit{keV}} \text{.}
\end{equation}

Note that the presence of magnetic fields can in principle strongly
alter the conductivity. Depending on the field configuration, it
can be suppressed in certain directions, or even in all directions in
cases of certain tangled fields. The field configuration in clusters
is not well understood, and it is currently debated to what degree
magnetic fields suppress conduction.  We will assume that the
modification of the conductivity can be expressed in terms of an
effective conductivity, which we parameterise as a fraction of the
Spitzer conductivity.

Even in the absence of magnetic fields, the Spitzer conductivity can
not be expected to apply down to arbitrarily low plasma densities.
Eventually, the scale length of the temperature gradient will become
comparable or smaller than the electron mean free path, at which point the
heat flux will saturate, with no further increase when the
temperature gradient is increased \citep{Cowie1977}.  This maximum heat
flux $j_{\rm sat}$ is given by 
\begin{equation}
  j_{\rm sat} \simeq 0.4\,n_{\rm e}k_{\rm B}T \left( \frac{2 k_{\rm
        B} T}{\pi m_{\rm e}} \right)^{1/2} \text{.}
\end{equation}
In order to have a smooth transition between the Spitzer regime and the
saturated regime, we limit the conductive heat flux by defining an
effective conductivity \citep{Sarazin1988} in the form
\begin{equation}
  \kappa_{\rm eff} = \frac{\kappa_{\rm sp}}{1 + 4.2\,
    {\lambda_{\rm e}}/{l_T}}.
  \label{kappa_sat}
\end{equation}
Here $l_T \equiv T / |\nablavec T|$ is the characteristic length-scale
of the temperature gradient.

\subsection{SPH formulation of conduction}

At first sight, equation~(\ref{eqn:conduction}) appears to be comparatively
easy to solve numerically. After all, the time evolution generated by the
diffusion equation smoothes out initial temperature
variations, suggesting that it should be quite `forgiving'
to noise in
the discretisation scheme, which should simply also be smoothed out.

In practice, however, there are two problems that make it surprisingly
difficult to obtain stable and robust implementations of the conduction
equation in cosmological codes. The first has to do with the
presence of second derivatives in equation~(\ref{eqn:conduction}),
which in
standard SPH kernel-interpolants can be noisy and sensitive to particle
disorder.  The second is that an explicit time integration method
can easily lead to an unstable integration if large local gradients
arise due to noise. We will discuss our approaches to solve these two
problems in turn.

A simple discretisation of the conduction equation in SPH can be obtained
by first estimating the heat flux for each particle applying standard
kernel interpolation methods, and then estimating the divergence in a
second step. However, this method has been shown to be quite sensitive to
particle disorder \citep{Brookshaw85}, which can be traced to the effective
double-differentiation of the SPH-kernel.  In addition, this method has the
practical disadvantage that an intermediate result, the heat flux vectors,
need to be computed and stored in a separate SPH-loop.

It is hence advantageous to use a simpler SPH discretisation of the Laplace
operator, which should ideally involve only first order derivatives of the
smoothing kernel.
Such a discretisation has been proposed before
\citep{Brookshaw85,Mo92}, and we here give a brief derivation of it in
three dimensions.

For a well-behaved field $Y(\vec{x})$, we can consider a
Taylor-series approximation for $Y(\vec{x}_j)$ in the proximity of
$Y(\vec{x}_i)$, e.g.
\begin{eqnarray}
  Y(\vec{x}_j) - Y(\vec{x}_i)  & = &    \nablavec Y\biggr|_{\vec{x}_i}
  \cdot  (\vec{x}_j - \vec{x}_i) + \notag \\
  & & \frac{1}{2} \frac{\partial ^2 Y}{\partial x_s \partial
  x_k}\biggr|_{\vec{x}_i} ( \vec{x}_j - \vec{x}_i )_s( \vec{x}_j -
  \vec{x}_i )_k +  \notag \\
  & & {\mathcal{O}}(\vec{x}_j-\vec{x}_i)^3  \text{.}
\end{eqnarray}
Neglecting terms of third and higher orders, we multiply this with
\begin{equation}
  \frac{(\vec{x}_j - \vec{x}_i) \nablavec_i W(\vec{x}_j - \vec{x}_i)}
  {|\vec{x}_j - \vec{x}_i|^2} \text{,}
  \label{eqn:multiply}
\end{equation} 
where  $W(\vec{x})=W(|\vec{x}|)$ is the
SPH smoothing kernel. Note that we choose this kernel to be
spherically symmetric and normalised to unity. The
expression in equation (\ref{eqn:multiply}) is well behaved for $\vec{x}_j \to
\vec{x}_i$ under these conditions. Introducing the notations
$\vec{x}_{ij} = \vec{x}_j - \vec{x}_i$ and $W_{ij} =
W(\vec{x}_j - \vec{x}_i)$, we integrate over all $\vec{x}_j$ and note that
\begin{equation}
  \int  \vec{x}_{ij}  
  \frac{\vec{x}_{ij} \nablavec_i W_{ij}}
  {|\vec{x}_{ij}|^2}   d^3{\vec{x}_j} = 0 ,
\end{equation}
\begin{equation}
  \int  (\vec{x}_{ij})_s (\vec{x}_{ij})_k 
  \frac{\vec{x}_{ij} \nablavec_i W_{ij}}
  {|\vec{x}_{ij}|^2}   d^3{\vec{x}_j} = \delta_{sk}.
\end{equation}

So the term linear in $\nablavec Y$ drops out, and the terms involving
off-diagonal elements of the Hesse matrix of $Y$ vanish, so that the
sum over the second order term simply reduces to $\nablavec^2 Y$. We
hence end up with
\begin{equation}
  \nablavec^2 Y \biggr|_{\vec{x}_i} = - 2 \int \frac{Y(\vec{x}_j) -
    Y(\vec{x}_i)}{| \vec{x}_{ij}|^2} \vec{x}_{ij} \nablavec_i W_{ij}
  \,{\rm d}^3{\vec{x}_j}  \text{.}
\end{equation}

This analytical approximation of the Laplacian can now be easily
translated into an SPH kernel interpolant.  To this end, we can
replace the integral by a sum over all particles indexed by $j$, and
substitute the volume element $\diff{^3\vec{x_j}}$ by its discrete SPH
analogue ${m_j}/{\rho_j}$. The values of the field $Y(\vec{x})$ at the
particle coordinates can be either taken as the value of the intrinsic
particle property that is evolved, $Y(\vec{x}_i) = Y_i$, or as a
kernel interpolant of these values, $Y(\vec{x}_i) =
\left< Y_i \right>$, where for example
\be \left< Y_i \right> = \sum_j Y_j
\frac{m_j}{\rho_j} W(\vec{x}_{ij}) . \ee
We then end up with a 
discrete SPH approximation of the Laplace operator in the form
\begin{equation}
  \nablavec^2 Y \biggr|_{i} = -2 \sum_j \frac{m_j}{\rho_j}  \frac{Y_j
    -Y_i}{|\vec{x}_{ij}|^2} \vec{x}_{ij}  \nablavec_i W_{ij} .
  \label{eqn:discrete}
\end{equation}

We now consider how this can be applied to the thermal conduction
problem, where the conductivity may also show a spatial variation.
Using the identity
\begin{equation}
  \nablavec \left( \kappa \nablavec T \right)
  = \frac{1}{2} \left[ \nablavec^2 ( \kappa T ) - T
    \nablavec^2  \kappa  + \kappa \nablavec^2 T \right],
\end{equation}
we can use our result from equation (\ref{eqn:discrete}) to write down a
discretised form of equation (\ref{eqn:conduction}):
\begin{equation}
  \frac{\diff{u}_i}{\diff{t}} =  \sum_j
  \frac{m_j}{\rho_i \rho_j} \frac{(\kappa_j + \kappa_i )\ (T_j -
    T_i)}{|\vec{x}_{ij}|^2} \vec{x}_{ij} \nablavec_i W_{ij} \text{.}
  \label{eqn:conduction_grad}
\end{equation}

This form is antisymmetric in the particles $i$ and $j$, and the
energy exchange is always balanced on a pairwise basis,
i.e.~conservation of thermal energy is manifest. Also, it is easy to
see that the total entropy can only increase,
and that heat always flows from higher to lower temperature.

The conductivities $\kappa_i$ and $\kappa_j$ in
equation~(\ref{eqn:conduction_grad}) are effectively arithmetically
averaged. \citet{Cleary99} proposed to make the replacement
\begin{equation}
   \frac{\kappa_i + \kappa_j}{2}\; \; \mapsto \; \;
  \kappa_{ij}= \frac{2\kappa_i \kappa_j}{\kappa_i + \kappa_j} \text{.}
\end{equation}
They showed that this ensures a continuous heat flux even in cases when the
heat conductivity exhibits a discontinuity, as for example along the
interface between different phases.  It is clear then that this
modification should also behave better when the conductivity changes
extremely rapidly on small scales, as it can happen for example in ICM gas
when cool particles get into direct contact with comparatively hot
neighbours. Indeed, we found this symmetrisation to give numerically more
robust behaviour, particularly in simulations that in addition to heat
conduction also followed radiative cooling.  Note that since we have
$\min(\kappa_i,\kappa_j)\le 2 \kappa_i \kappa_j / (\kappa_i + \kappa_j) \le
2 \min(\kappa_i,\kappa_j)$, the \citeauthor{Cleary99} average stays always
close to the smaller of the two conductivities involved, to within a factor
of two.

\subsection{Numerical implementation details}

We have implemented heat conduction in a new version of the massively
parallel TreeSPH-code {\small GADGET} \citep{SprGadget2000}, which is a
general purpose code for cosmological structure formation. Unlike earlier
public releases of the code, the present version, {\small
  GADGET-2}, uses the `entropy formulation' of SPH proposed by
\cite{SprHe01} which conserves both energy and entropy (in regions without
shocks) for fully adaptive smoothing lengths.  In this formulation, an
entropic function \be A = (\gamma-1)\frac{u}{\rho^{\gamma-1}} \ee is
evolved as independent variable for each particle, instead of the thermal
energy per unit mass. Note that $A=A(s)$ is only a function of the
thermodynamic entropy  $s$ per unit mass. 

For consistency with this formalism, we need to express the heat conduction
equation in terms of entropy. This is easily accomplished in an isochoric
approximation, noting that the temperature can be expressed as $T = \mu/k_B
\ A \rho^{\gamma-1}$, where $\mu$ is the mean molecular weight.  This
results in
\begin{equation}
\frac{\diff{A}_i}{\diff{t}} =
\frac{2 \mu}{k_B} 
\frac{\gamma-1}{\rho_i^{\gamma-1}}
\sum_j  \frac{m_j \kappa_{ij}}{\rho_i \rho_j} 
\left( \frac{A_j}{\rho_j^{1-\gamma}}  -  \frac{A_i}{\rho_i^{1-\gamma}}\right)
\frac { \vec{x}_{ij} \nablavec_i
W_{ij}}{|\vec{x}_{ij}|^2} \text{.}
\label{eqn:conduction_grad2}
\end{equation}

One problematic aspect of the heat conduction equation is that small-scale
numerical noise in the temperature field can generate comparatively large
heat flows, simply because this noise can involve small-scale gradients of
sizable magnitude. Since we are using an explicit time integration scheme
for the hydrodynamical evolution, this immediately raises the danger of
instabilities in the integration. Unless extremely small timesteps or an
implicit integration scheme are used, the energy exchange between two
particles due to a small-scale temperature difference can become so large
that the explicit time integration ``overshoots'', thereby potentially {\em
  reversing} the sign of the temperature difference between the two
particles in conductive contact. This is not only incorrect, but makes
it possible for the temperature differences to grow quickly in an oscillatory
fashion, causing an instable behaviour of the integration.

We have found that a good method to avoid this problem is to use a kernel
interpolant for the temperature field (or entropy field) in the
discretisation of the heat conduction equation
(\ref{eqn:conduction_grad2}), instead of the individual particle
temperature values themselves. The interpolant represents a smoothed
version of the noisy sampling of the temperature field provided by the
particle values, so that on the scale of the SPH smoothing length,
small-scale noise in the heat flux is strongly suppressed.  Particles will
still try to equilibrate their temperatures even within the smoothing
radius of each particle, but this will happen at a damped rate.
Heat-conduction due to temperature gradients on larger scales is unaffected
however.  In a later section of this work, we will explicitly demonstrate
how this improves the stability of the time integration, particularly
when individual and adaptive timesteps are used.

For definiteness, the interpolant we use is a smoothed version
$\overline{A}_i$ of the entropy, defined by
\begin{equation}
\rho_i^{\gamma} \overline{A}_i = \sum_j m_j \rho_j^{(\gamma - 1)} A_j  W_{ij} \text{.}
\label{eqn:smoothing}
\end{equation}
We then replace $A_i$ and/or $A_j$ on the right-hand-side of equation
(\ref{eqn:conduction_grad2}) with the interpolants $\overline{A}_i$ and
$\overline{A}_j$.  Note that the weighting by $\rho^{(\gamma-1)}$ ensures
that we obtain a value of $\overline{A}$ that corresponds to a smoothed
temperature field, as required since conduction is driven by gradients in
temperature and not entropy.  However, since the density values need to be
already known to evaluate this interpolant, this unfortunately requires an
additional SPH loop, which causes quite a bit of computational overhead.
This can in principle be avoided if the weighting with densities in
equation (\ref{eqn:smoothing}) is dropped, which according to our tests
appears to be sufficiently accurate in most situations.

Note that although we use the SPH interpolant for the entropy values
in equation~(\ref{eqn:conduction_grad2}), we still
compute the values for the particle conductivity $\kappa$ based
on the intrinsic particle temperature and not on its smoothed counterpart.

While the above formulation manifestly conserves thermal energy, this
property may get lost when individual and adaptive time steps are used,
where for a given system step only a subset of the particles is
evolved. There can then be particle pairs where only one particle is active
in the current system step, while the other is not, so that a `one-sided'
heat conduction may occur that causes a (brief) violation of energy
conservation during the step.  While most of the resulting energy imbalance
will only be a temporary fluctuation that will be compensated as soon as
the `inactive' particle in the pair is evolved, the very strong temperature
dependence of the conductivity may produce sizable errors in this
situation, particularly when coarse timestepping is used. We therefore
decided to implement an explicitly conservative scheme for the heat
exchange, even when adaptive and individual timesteps are used.

\begin{figure*}
  \begin{center}
    \includegraphics[width=5.6cm]{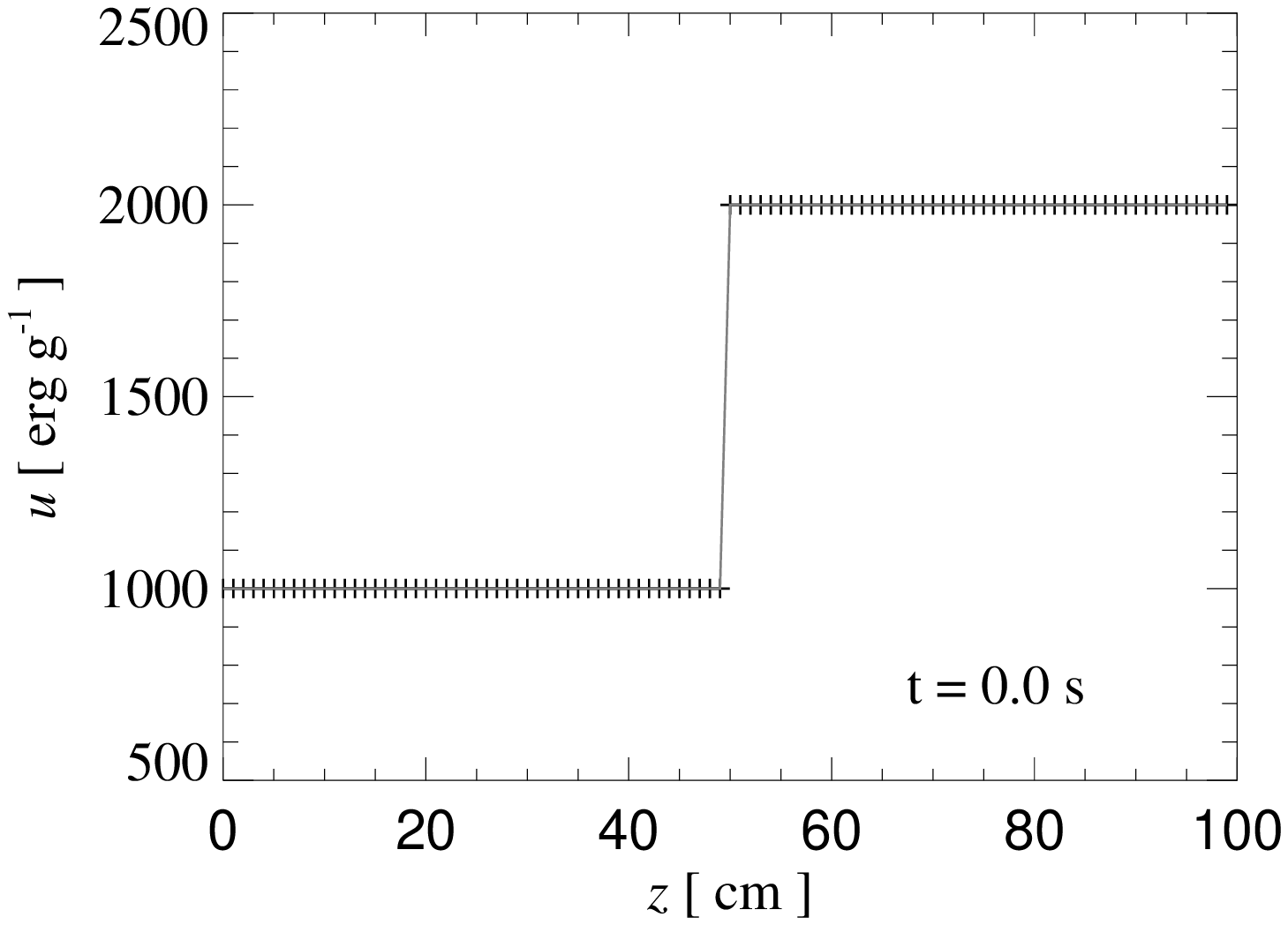}
    \includegraphics[width=5.6cm]{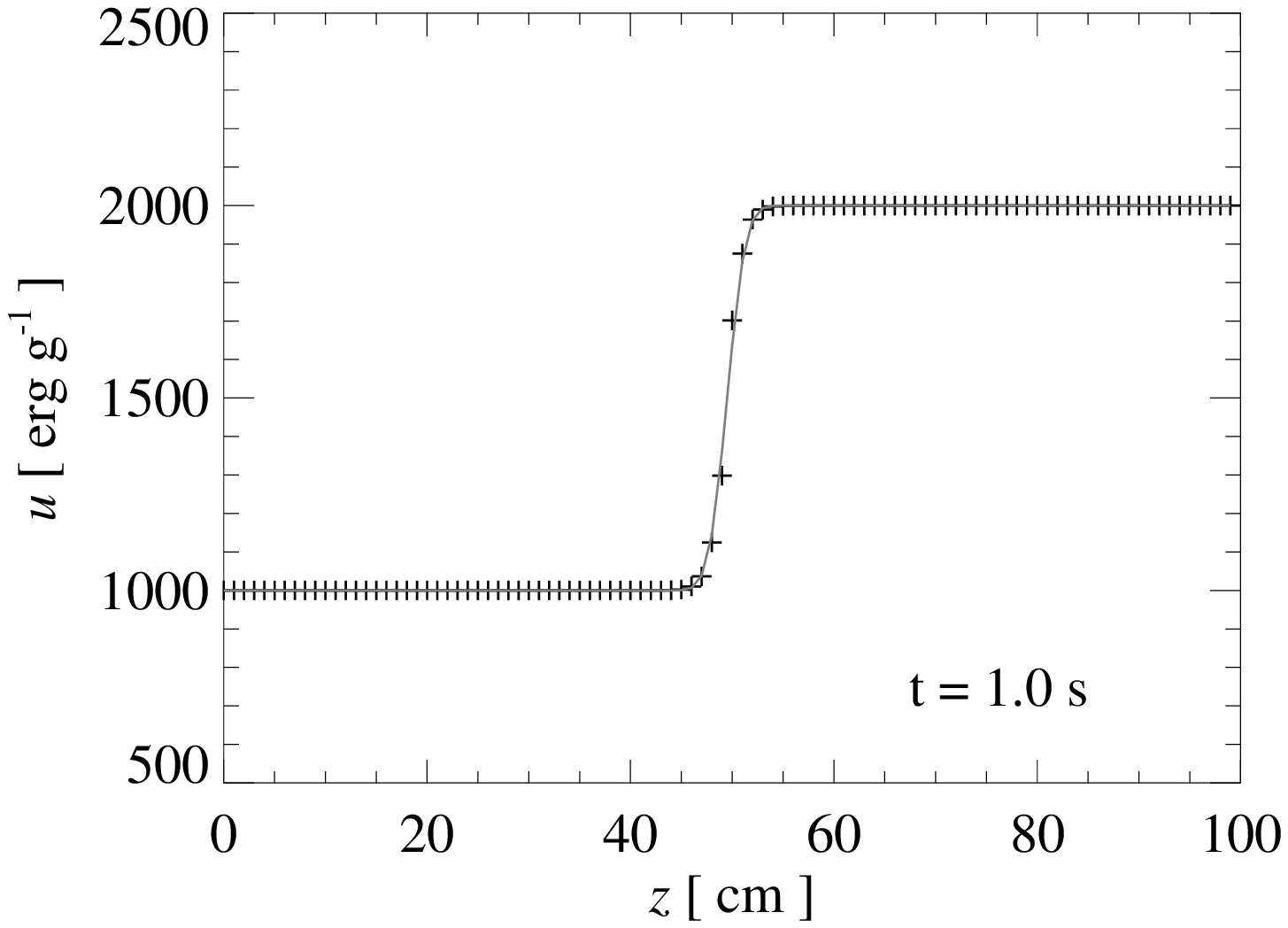}
    \includegraphics[width=5.6cm]{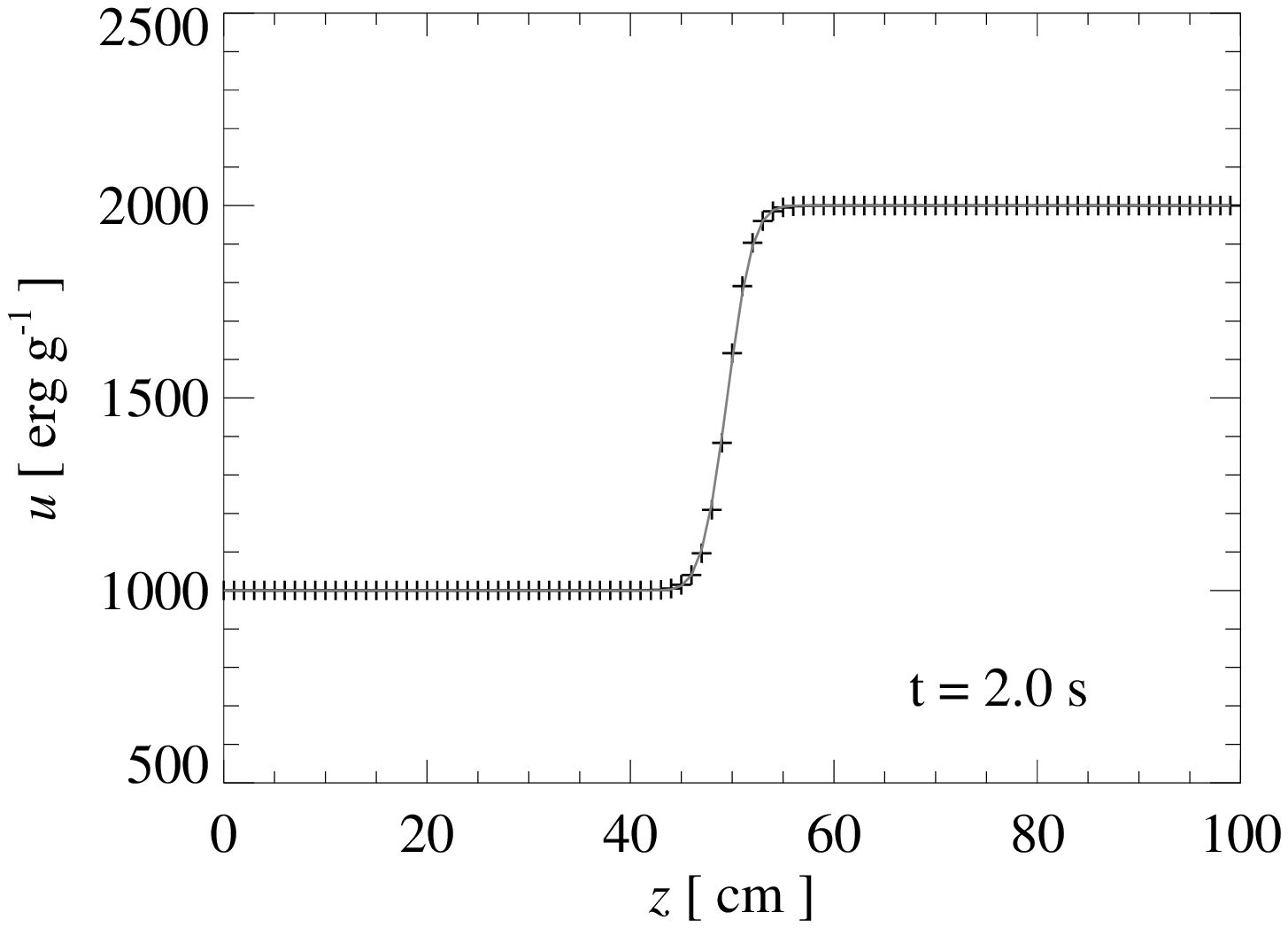}
    \includegraphics[width=5.6cm]{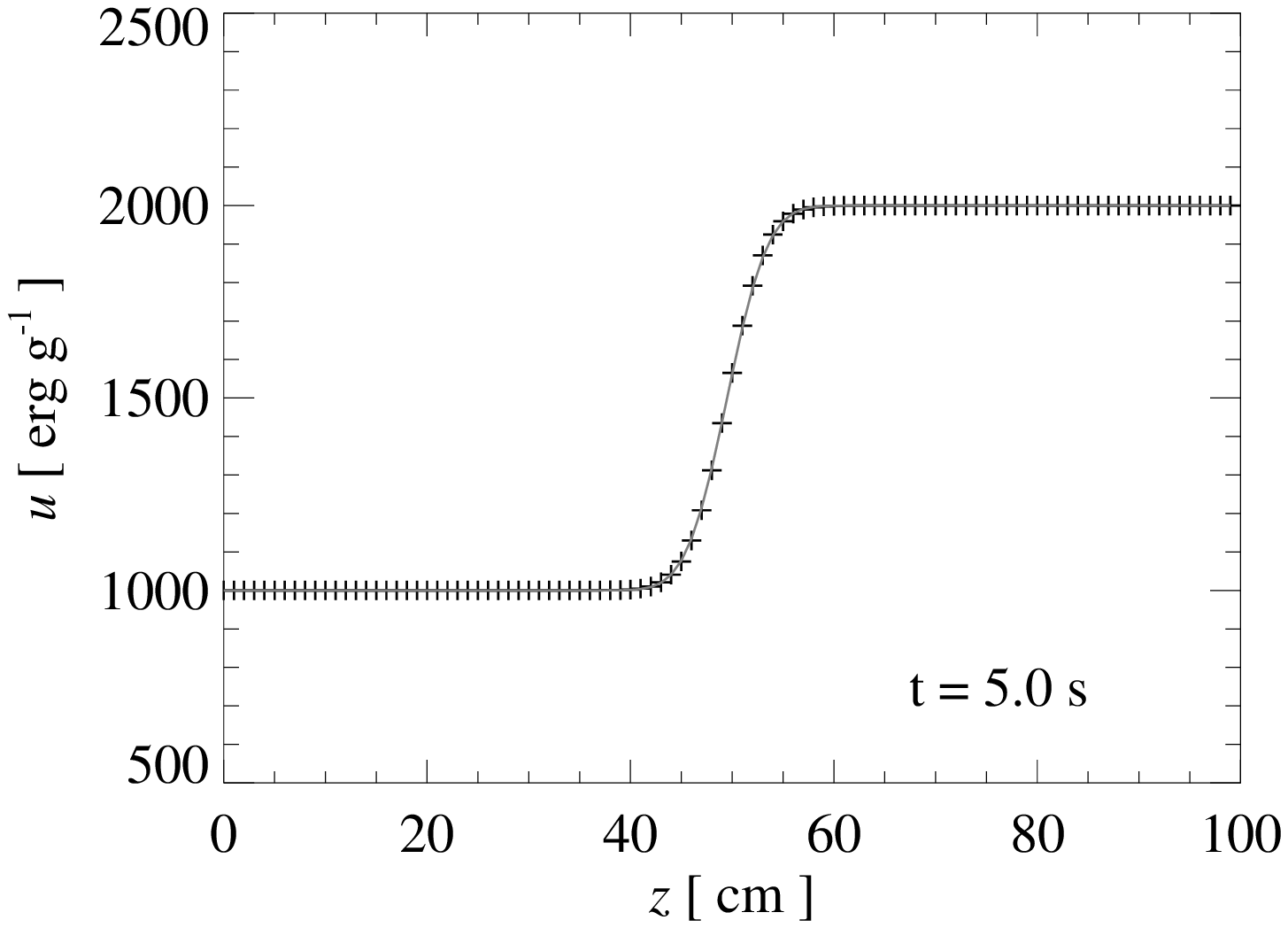}
    \includegraphics[width=5.6cm]{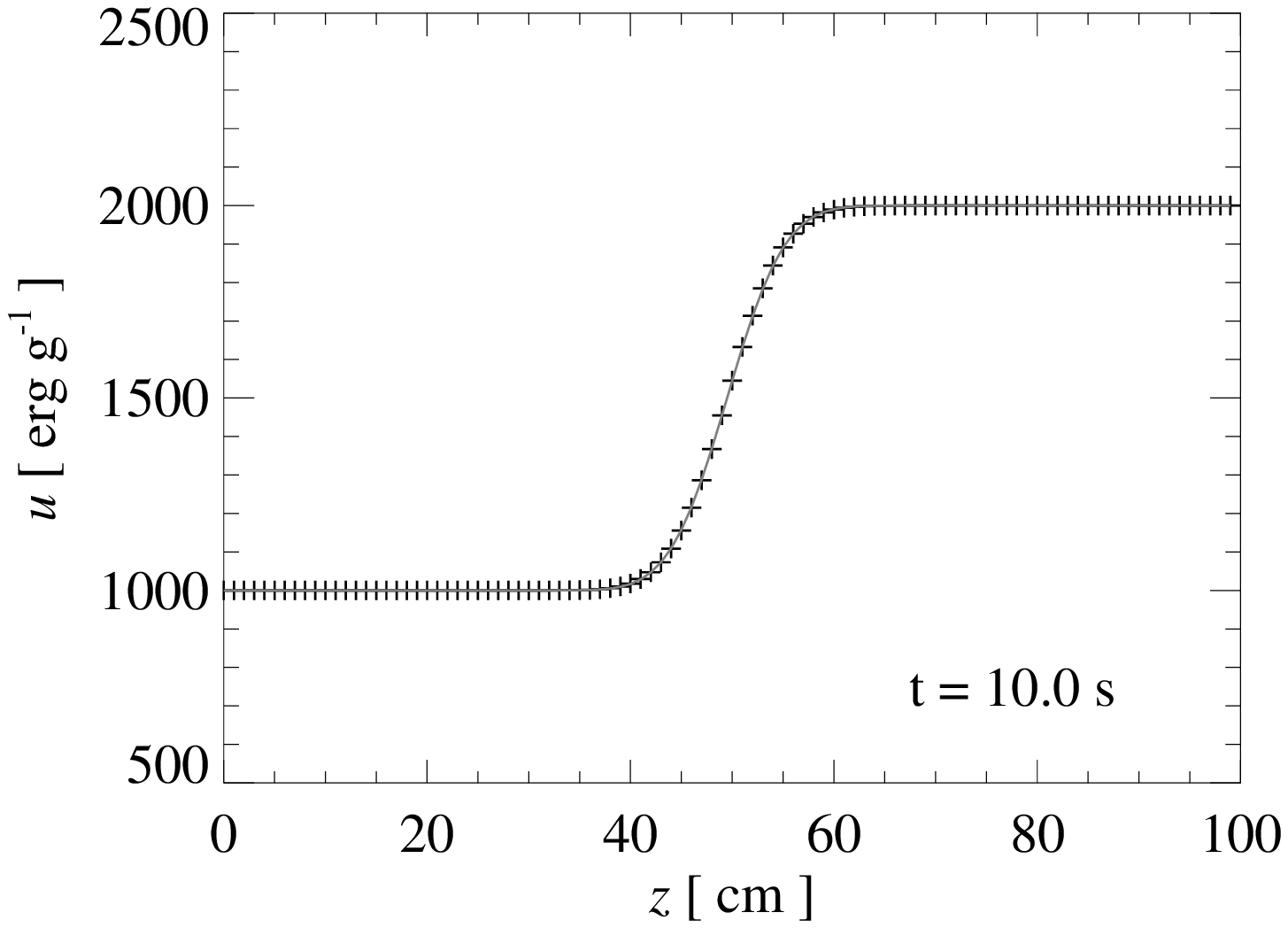}
    \includegraphics[width=5.6cm]{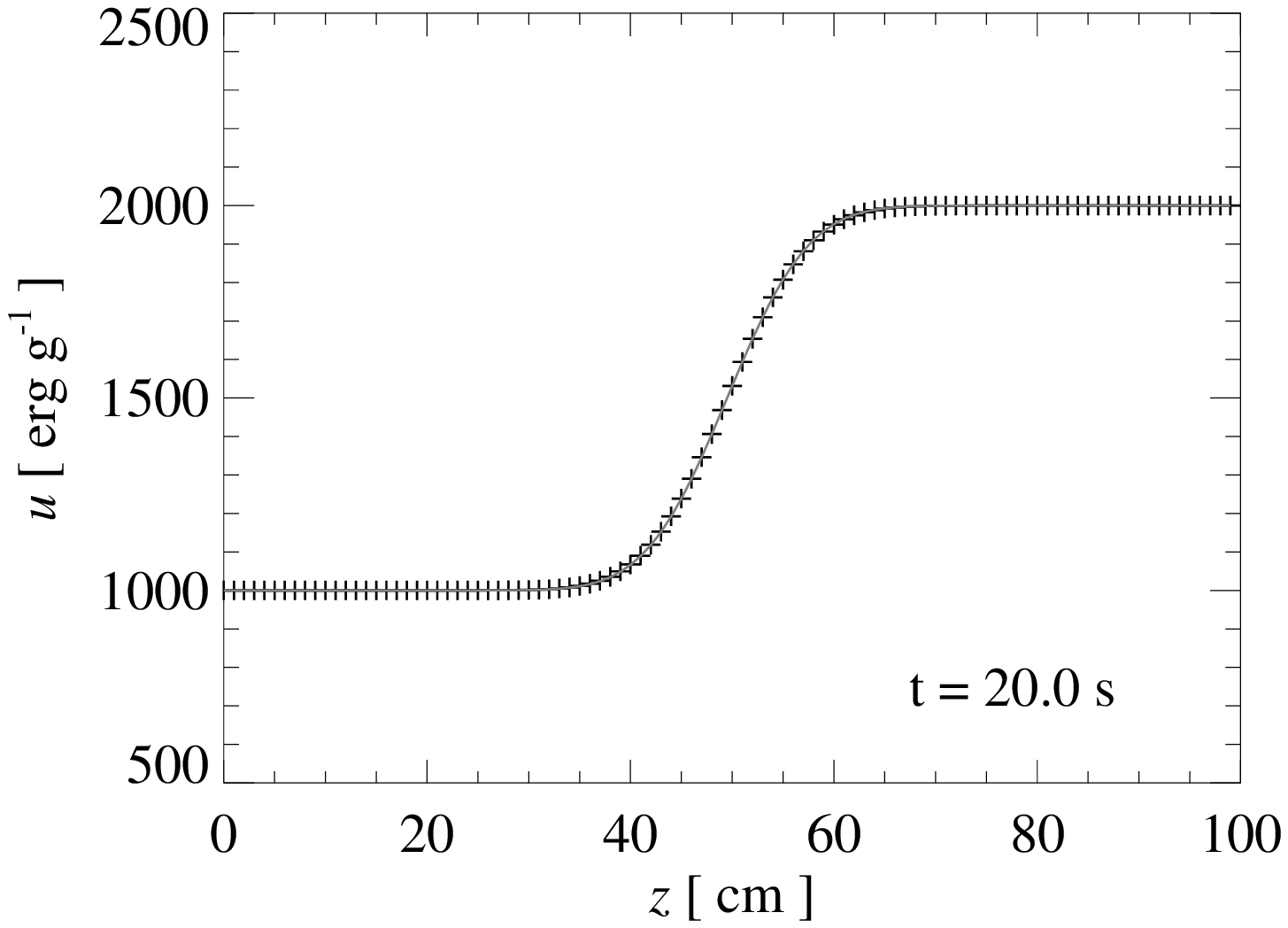}
  \end{center}
  \caption{Time evolution of the temperature profile of two slabs of
    solid material, brought in contact with each other at $t = 0$ along
    the $z= 50\,{\rm cm}$ plane, with an initial difference of thermal
    energy per unit mass of $\Delta u = 1000\,{\rm erg\,g^{-1}}$. Crosses
    mark numerical results, and the solid line is the analytic solution of
    the heat conduction equation.
    \label{figSlab}  }
\end{figure*}

To this end, we define a pairwise exchange of heat energy as \be
E_{ij} \equiv \frac{2 \mu}{k_B} \frac{m_i m_j \kappa_{ij}}{\rho_i
\rho_j} \left( \frac{A_j}{\rho_j^{1-\gamma}} -
\frac{A_i}{\rho_i^{1-\gamma}}\right) \frac { \vec{x}_{ij} \nablavec_i
W_{ij}}{|\vec{x}_{ij}|^2} \text{.} \label{eqn:pairenergy}  \ee A simple translation 
into a finite difference scheme for the time evolution
would then be \be m_i u_i' = m_i u_i + \Delta t_i \sum_j E_{ij},
\label{timeint1}\ee 
where $u_i'$ and $u_i$ can also be expressed in terms of the corresponding
entropy values $A_i'$ and $A_i$.  However, for individual and variable
timesteps, only a subfraction of particles will be `active' in the current
system timestep. These particles have individual timesteps $\Delta t_i$,
while the `inactive' particles can be formally assigned $\Delta t_i=0$ for
the step. Equation~(\ref{timeint1}) then clearly does not guarantee
detailed energy conservation in each system step.

We recover this property in the following way. We update the energy of
particles according to \be m_i u_i' = m_i u_i + \frac{1}{2} \sum_{jk}
\Delta t_j (\delta_{ij}-\delta_{ik}) E_{jk}\label{timeint2} \ee in
each system step. In practice, the double-sum on the right hand side
can be simply computed using the usual SPH loop for active
particles. Each interacting pair found in the neighbour search for
active particle $i$ is simply used to change the thermal energy of
particle $i$ by $\Delta t_i E_{ij}/2$, and that of the neighbouring
particle $j$ by $-\Delta t_i E_{ij}/2$ as well. Note that if particle
$j$ is active, it carries out a neighbour search itself and will find
$i$ itself, so that the total energy change of $i$ due to the presence
of $j$ is given by $(\Delta t_i + \Delta t_j)E_{ij}/2$. 
Energy is conserved by construction in this scheme, independent of the
values if individual timesteps of particles. If all particles have
equal steps, equation (\ref{timeint2}) is identical to the form of
(\ref{timeint1}). As before, in our SPH implementation we convert the
final thermal energy change to a corresponding entropy change. These
entropy changes are applied instantaneously for all particles at the
end of one system step, so that even particles that are inactive at
the current system step may have their entropy changed by
conduction. The resulting low order of the time integration scheme is
sufficient for the diffusion equation.

If conduction is so strong that the timescale of conductive
redistribution of thermal energy is short compared to relevant
dynamical timescales of the gas, the usual hydrodynamic timestep
selected by the code based on the Courant-criterion may become too
coarse to follow conduction accurately.  We therefore introduce an
additional timestep criterion in simulations that follow thermal
conduction. To this end, we limit the maximum allowed timestep to some
prescribed fraction of the conduction time scale, ${A}/{|\dot{A}_{\rm
cond}|}$, namely
\begin{equation}
  \Delta t_{\rm cond} =  \alpha \times \frac{A}{|\dot{A}_{\rm cond}|} \text{,}
  \label{eqn:timestep}
\end{equation}
where $\alpha$ is a dimensionless accuracy parameter.  In our
simulations, we typically employed a value of $\alpha = 0.25$, which
has provided good enough accuracy at a moderate cost of CPU time.

Finally, in order to account for a possible limitation of conduction by
saturation, we compute the gradient of the (smoothed) temperature field in
the usual SPH fashion, and then use it to compute $l_T$ and the
saturation-limited conductivity based on equation~(\ref{kappa_sat}).

\section{Illustrative test problems}

In this section, we verify our numerical implementation of thermal
conduction with a number of simple test problems that have known
analytic solutions. We will also investigate the robustness of our
formulation with respect to particle disorder, and (initial) noise in
the temperature field.

\subsection{Conduction in a one-dimensional problem}

\begin{figure*}
  \begin{center}
    \includegraphics[width=7.5cm]{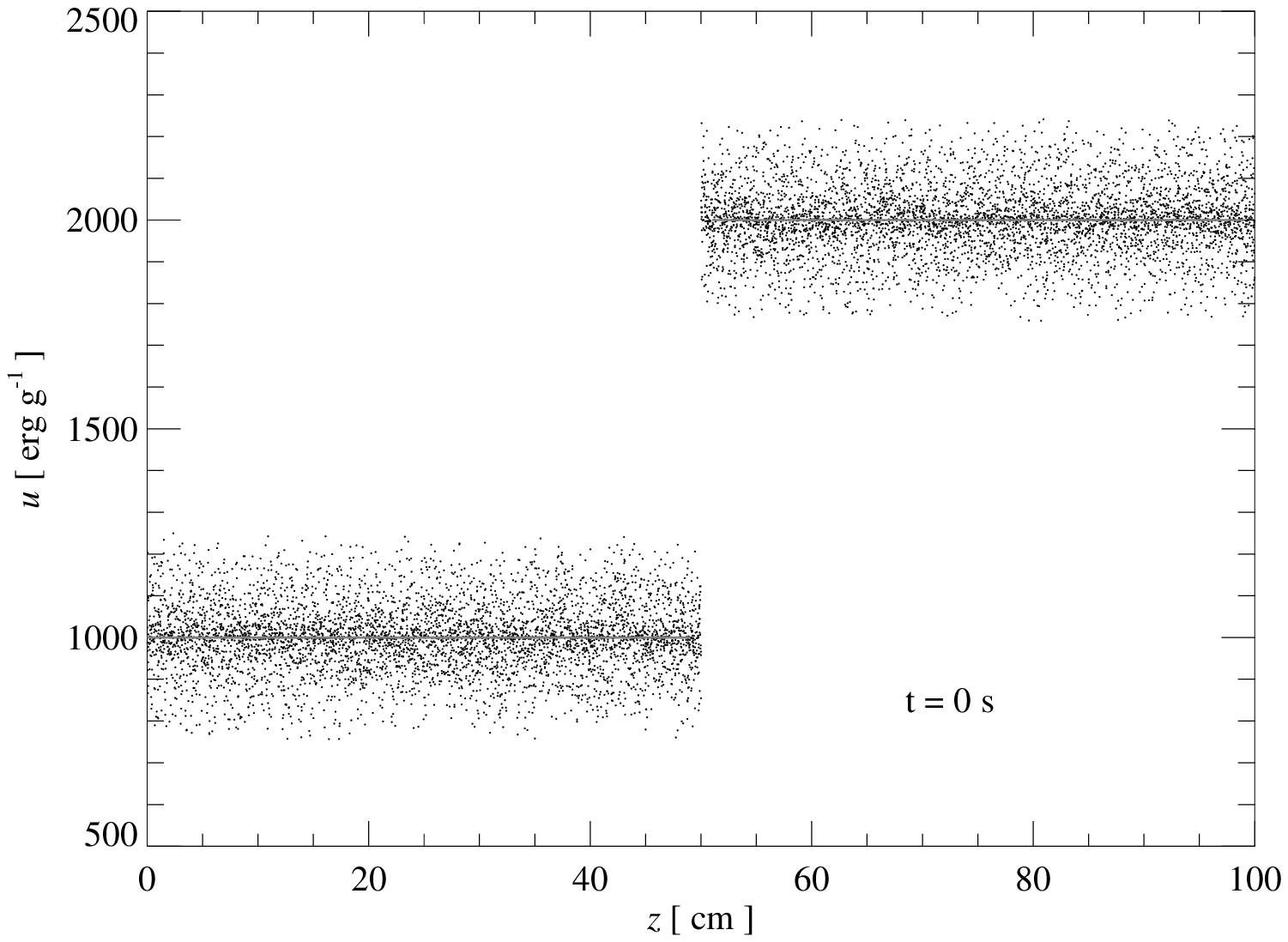}
    \includegraphics[width=7.5cm]{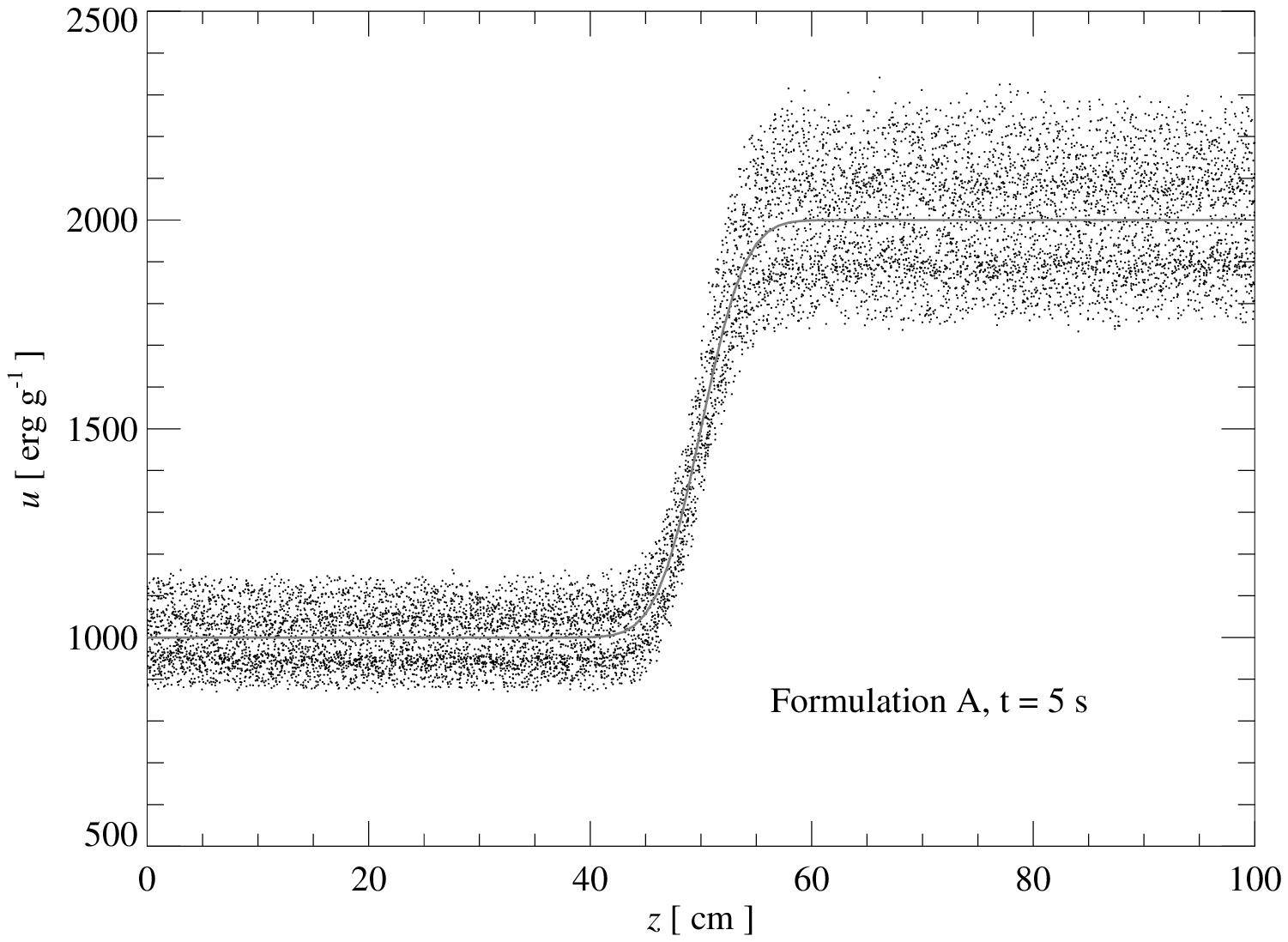}
    \includegraphics[width=7.5cm]{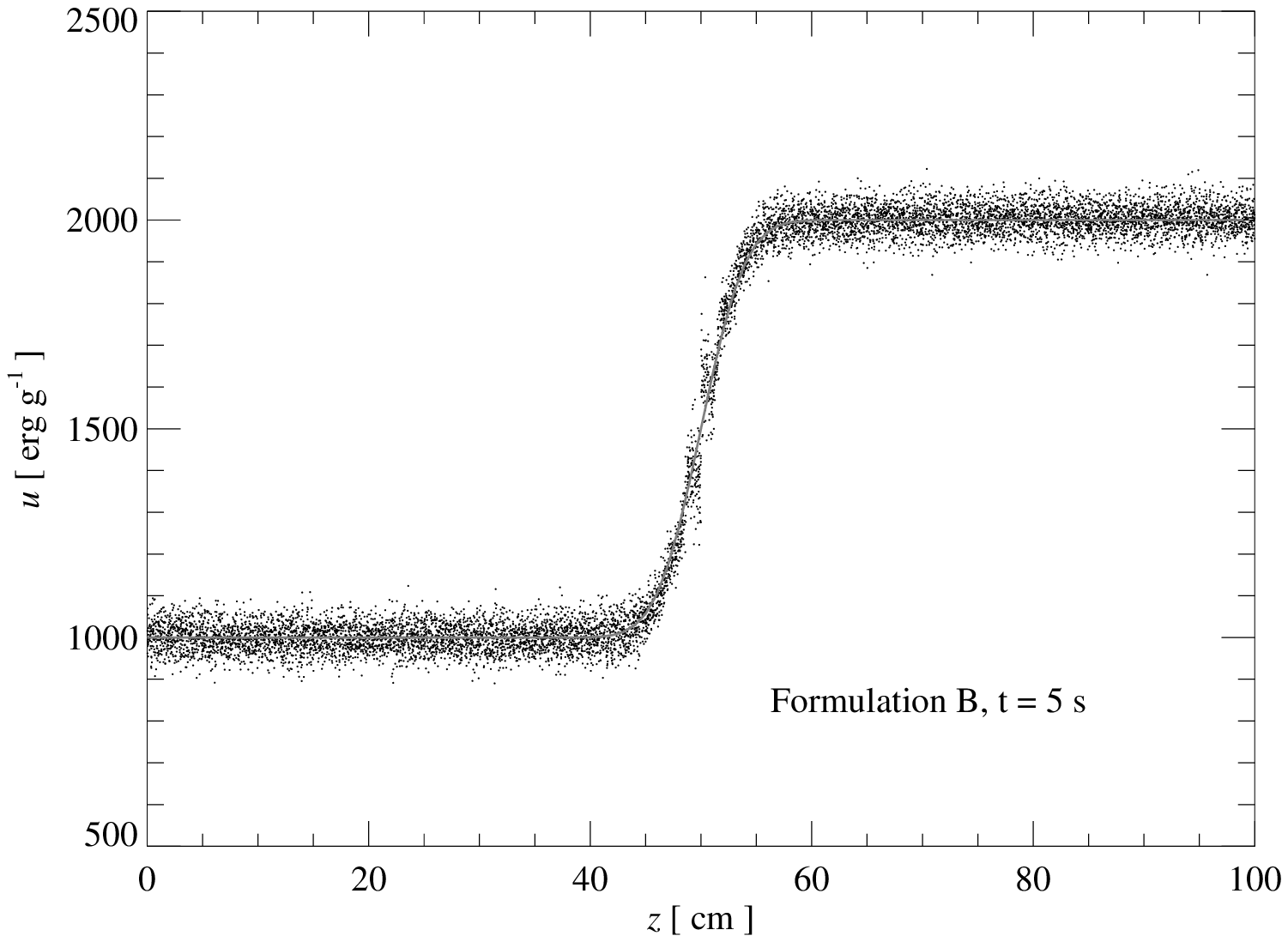}
    \includegraphics[width=7.5cm]{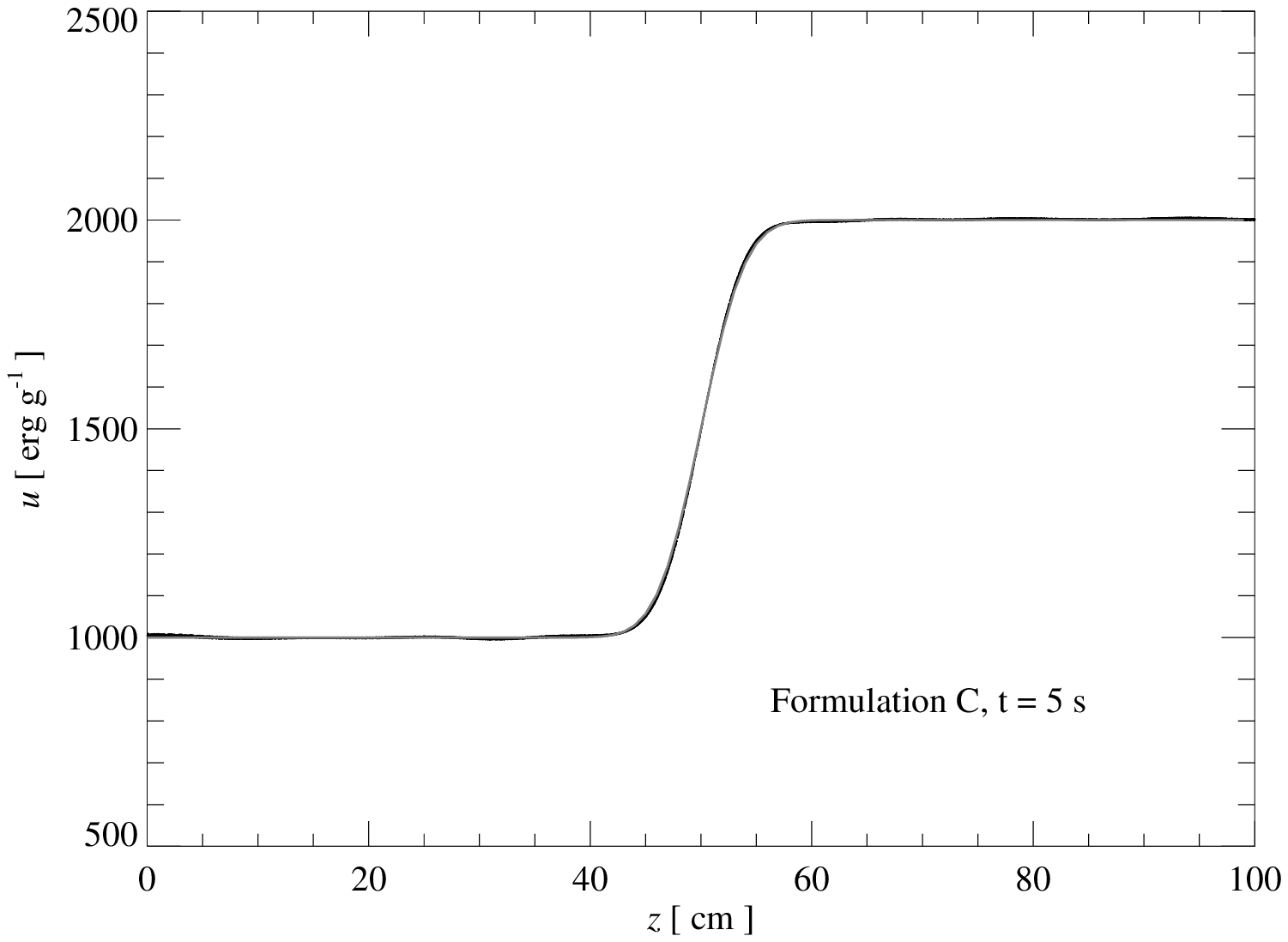}
  \end{center}
  \caption{Comparison of the numerical stability of the SPH
    discretisation scheme when different formulations for the conduction
    equation (\ref{eqn:pairenergy}) are used. The top left panel shows the
    initial conditions for the simple conduction problem considered in
    Fig.~\ref{figSlab}, but perturbed with artificial Gaussian noise in
    the temperature field. The top right panel shows the evolved state
    after $5\,{\rm sec}$ when formulation (A) is employed, where the
    individual particle temperatures are used directly. The lower left
    panel compares this with formulation (B), where the temperatures on
    the right hand-side of equation (\ref{eqn:pairenergy}) are taken to be
    kernel interpolants of the temperature field. Finally, the bottom
    right panel gives the result for formulation (C), which represents a
    mixed scheme that uses both the individual temperatures of particles,
    and the smoothed temperature field. This scheme proves to be the most
    robust against local noise, which is quickly damped
    away.\label{fig:smoothing_compare}}
\end{figure*}

We first consider two solid slabs with different initial temperatures,
brought into contact with each other at time $t=0$.  The slabs were
realized as a lattice of SPH particles, with dimensions
$100\times10\times10$, and an equidistant particle-spacing of $1\
\unit{cm}$. To avoid perturbations of the 1D-symmetry, the simulation
volume was taken to be periodic along the two short axes, while kept
non-periodic along the long axis, allowing us to study surface
effects, if present, on the sides of the slabs that are not in
contact. Note that the test was carried out with our 3D code.
All particle velocities were set to zero initially, and kept at this
value to mimic a solid body. The thermal conductivity was set to a
value corresponding to
$\alpha = {\kappa}/{(c_v \rho)} = 1\ \unit{cm}^2\,\unit{s}^{-1}$ 
throughout both materials, independent of temperature.

In Figure~\ref{figSlab}, we compare the time evolution of our
numerical results for this test with the analytical solution of the
same problem, which is given by
\begin{equation}
u(z,t) = u_0 + \frac{\Delta u}{2} \, {\rm erf} \left( \frac{z-z_m}{\sqrt{4 \alpha
      t}} \right),
\end{equation}
where $z_m$ gives the position of the
initial difference of size $\Delta u$ in thermal energy, and $u_0$ is
the mean thermal energy.  We see that the numerical solution tracks
the analytic result very nicely.

We have also repeated the test for a particle configuration
corresponding to a `glass' \citep{White96}, with equally good results.
For a Poisson distribution of particles, we noted however a small
reduction in the speed of conduction when a small number of neighbours
of 32 is used. Apparently, here the large density fluctuations due to
the Poisson process combined with the smallness of the number of
neighbours leads to somewhat poor coupling between the particles. This
effect goes away for larger numbers of neighbours, as expected. Note
however that in practice, due to the pressure forces in a gas, the
typical configuration of tracer particles is much more akin to a glass
than to a Poisson distribution.

Next, we examine how robust our formulation is with respect to
small-scale noise in the temperature field.  To this end, we repeat the
above test using the glass configuration, but we perturb the initial
thermal energies randomly with fluctuations at an rms-level of $\sigma =
80\ \unit{erg g}^{-1}$.  Further, we increase the maximum timestep allowed
for particles to $\Delta t = 1.0\ \unit{s}$. 

We discussed previously that we can either use the particle values of the
temperatures (or entropy) in the right-hand-side of equation
(\ref{eqn:pairenergy}), or kernel interpolants thereof. Here we compare the
following different choices with each other:
\begin{enumerate}
\item[(A)] Basic formulation: Use particle values $A_i$ and $A_j$.
\item[(B)] Smoothed formulation: Use $\overline{A}_i$ and $\overline{A}_j$.
\item[(C)] Mixed formulation: Use $A_i$ and $\overline{A}_j$.
\end{enumerate}
The mixed formulation (C) may at first seem problematic, because its
pair-wise antisymmetry is not manifest. However, since we use equation
(\ref{timeint2}) to exchange heat between particles, conservation of
thermal energy is ensured also in this case. When all particles have equal
timesteps, formulation (C) would correspond to an arithmetic average of (A)
and (B).

\begin{figure*}
  \includegraphics[width=7.5cm]{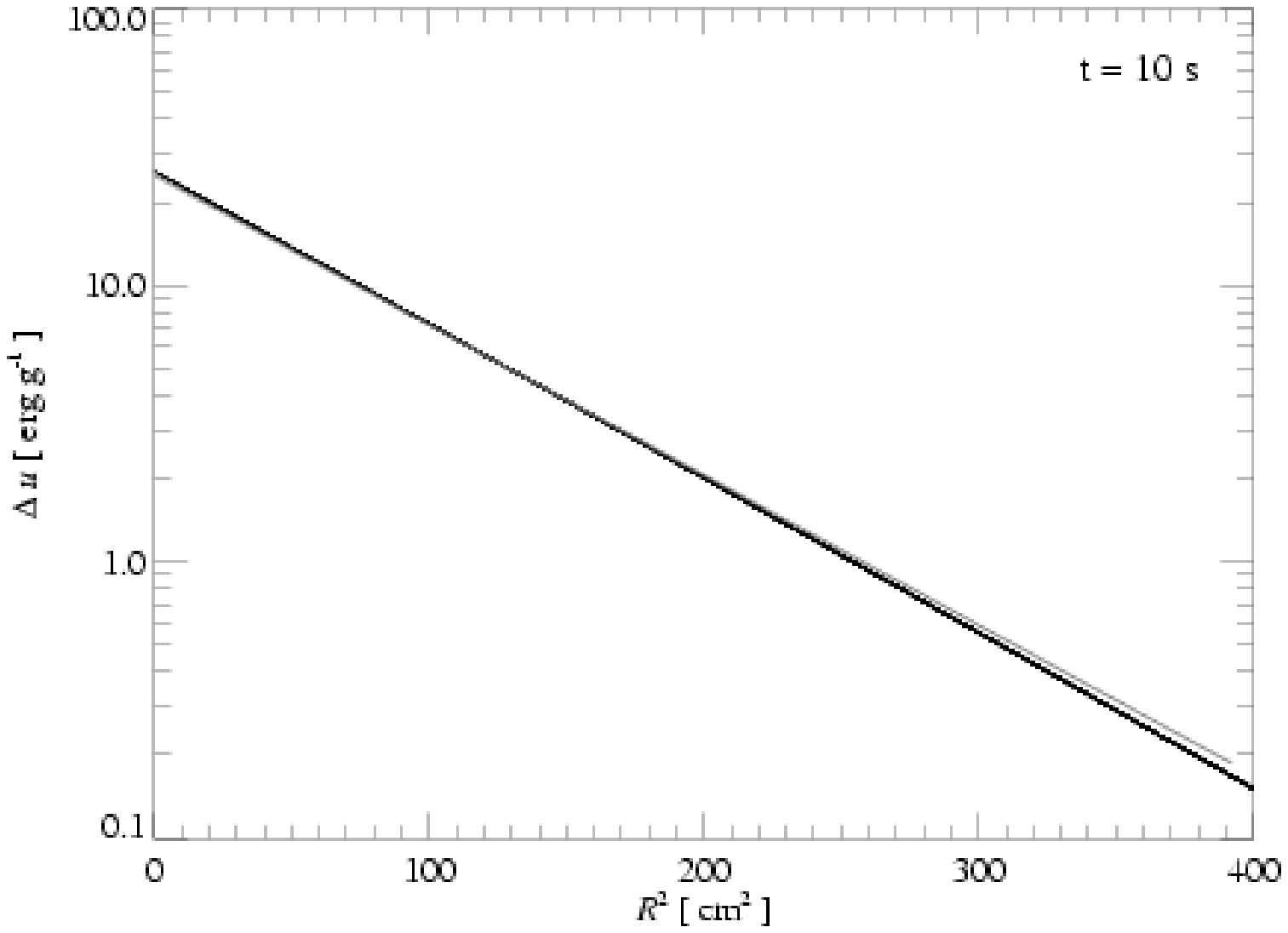}
  \includegraphics[width=7.5cm]{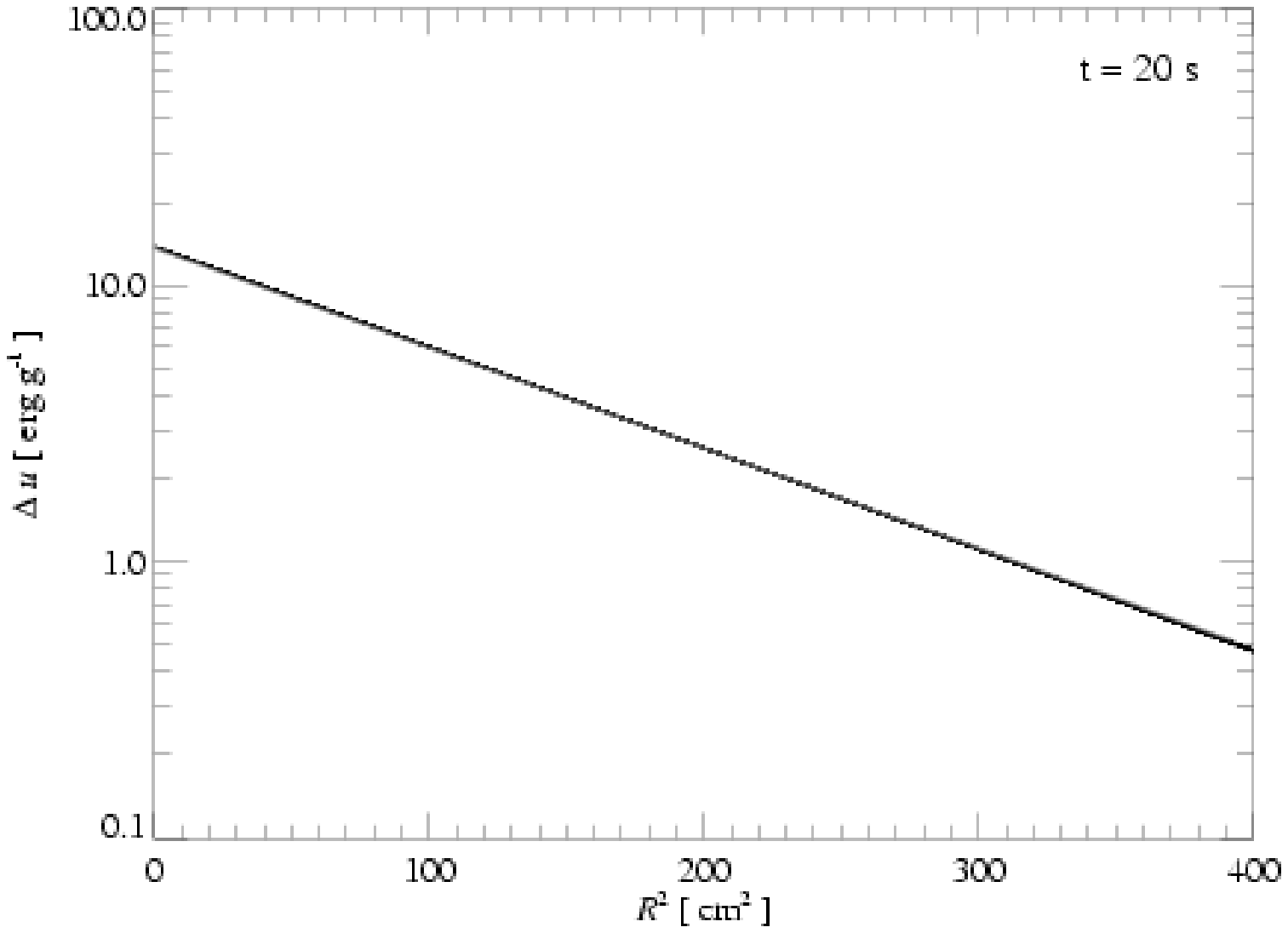}
  \includegraphics[width=7.5cm]{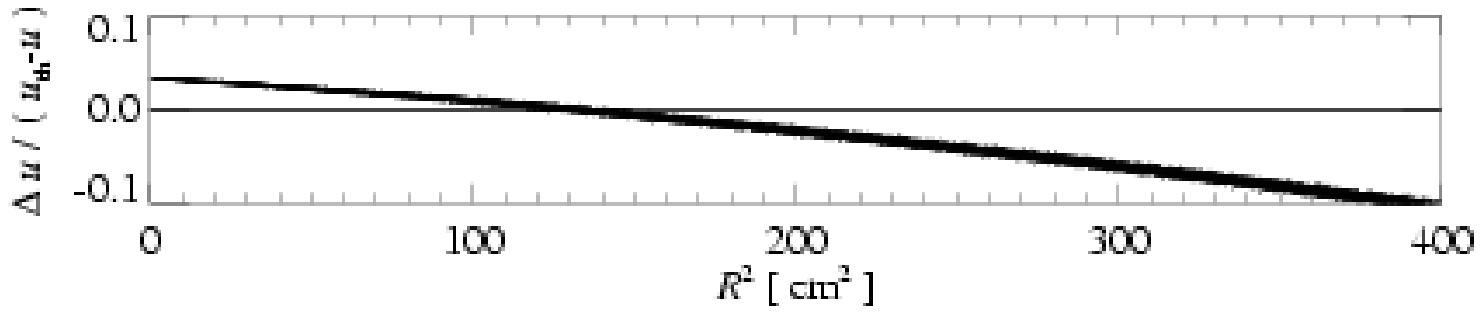}
  \includegraphics[width=7.5cm]{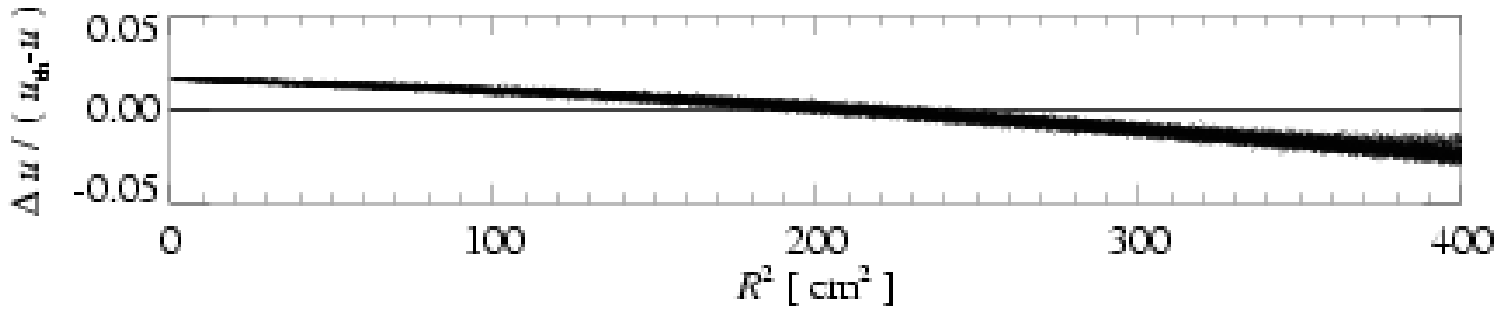}
  \caption{Time evolution of the temperature field in an elementary
    three-dimensional conduction problem, displayed at two different
    times. 
    We here consider the spreading
    of a narrow Gaussian temperature profile, which corresponds to the
    Green's function of the conduction problem for constant
    conductivity.  Dots show the specific energies of individual
    simulation particles, while the solid line marks the analytical
    result. Relative deviations are shown in the lower panels. The
    numerical result maintains the Gaussian profile very well at all
    times.  At early times, when the profile is sampled with few
    particles, a small reduction in the effective speed of conduction is
    seen, which however becomes increasingly unimportant at later times.}
  \label{fig:3DTest}
\end{figure*}

In Figure~\ref{fig:smoothing_compare}, we compare the result obtained for
these three formulations after a simulation time of 5 sec.  The
artificially perturbed initial conditions are shown in the top left panel.
Interestingly, while all three different formulations are able to recover
the analytic solution in the mean, they show qualitatively different
behaviour with respect to the imprinted noise. When the intrinsic particle
values for the temperatures are used (top right panel), very large
pair-wise gradients on small scales occur that induce large heat exchanges.
As a result, the particles oscillate around the mean, maintaining a certain
rms-scatter which does not reduce with time, i.e.~an efficient relaxation
to the medium temperature does not occur.  Note that the absolute size of
the scatter is larger in the hot part of the slab. This is a result of the
timestep criterion (\ref{eqn:timestep}), which manages to hedge the rms-noise
to something of order $\sim \alpha T$.  Reducing the timestep parameter
$\alpha$ can thus improve the behaviour, while simultaneously increasing
the computational cost significantly. For coarse timestepping (or high
conductivities) the integration with this method can easily become
unstable.

Formulation (B), which uses the smoothed kernel-interpolated temperature
field for both particles in each pair, does significantly better in this
respect (bottom left panel).  However, the particle temperatures only very
slowly approach the local mean value and hence the analytical solution in
this case. This is because the smoothing here is quite efficient in
eliminating the small-scale noise, meaning that a deviation of an
individual particle's temperature from the local mean is decaying only very
slowly.

The mixed formulation (C), shown in the bottom right, obviously shows the
best behaviour in this test. It suppresses noise quickly, matches the
analytical solution very well, and allows the largest timesteps of all
schemes we tested. In this formulation, particles which have a large
deviation from the local average temperature try to equalise this
difference quickly, while a particle that is already close to the mean is
not `pulled away' by neighbours that have large deviations. Apparently,
this leads to better behaviour than for schemes (A) and (B), particularly
when individual timesteps are used. We hence choose formulation (C) as our
default method.

\subsection{Conduction in a three-dimensional problem}

As a simple test of conduction in an intrinsically three-dimensional problem,
we consider the temporal evolution of a point-like
thermal energy perturbation. The time evolution
of an initial
$\delta$-function 
is given
by the three-dimensional Green's function for the conduction
problem,
\begin{equation}
  G( x, y, z, t ) = \frac{1}{(4 \pi \alpha t)^{3/2}} \exp \left( -
  \frac{x^2 + y^2 + z^2}{4 \alpha t} \right) \text{,}
  \label{eqn:3dGreen}
\end{equation}
where $c_v = u/T$ is the heat capacity.  We again consider a solid
material, realized with a glass configuration of $50^3$ simulation
particles with a mass of $1\,\unit{g}$ each and a mean particle
spacing of $1\,{\rm cm}$.  We choose a conductivity of $\kappa c_v =
1\,\unit{cm}^2\,\unit{s}^{-1}$, as before.  We give the material a
specific energy per unit mass of $u_0 = 1000\,\unit{erg}\
\unit{g}^{-1}$, and add a perturbation of $10000.0\
\unit{erg}\ \unit{cm}^3\,\unit{g}^{-1} \times G(x,y,z, t_0 = 10\
\unit{sec})$. These initial conditions correspond to a
$\delta$-function perturbation that has already evolved for a brief
period; this should largely eliminate timing offsets that would 
arise in the later evolution if we let the intrinsic SPH smoothing
wash out a true initial $\delta$-function.

Figure~\ref{fig:3DTest} shows the specific energy profile after evolving
the conduction problem for an additional 10 or 20 seconds, respectively.
In the two bottom panels, we show the relative deviation from the analytic
solution. Given that quite coarse timestepping was used for this problem
(about $\sim 32$ timesteps for 20 sec), the match between the theoretical
and numerical solutions is quite good. Even at a radial distance of
$20\,\unit{cm}$, where the Gaussian profile has dropped to less than one
hundredth of its central amplitude, relative deviations are at around 10
percent for $t = 20\ \unit{sec}$ and below 3 percent at $t = 30\ 
\unit{sec}$. Note however that at all times the numerical solution
maintains a nice Gaussian shape so that the deviations can be interpreted
as a small modification in the effective conductivity. We then see that at
late times, when the temperature gradients are resolved by more particles,
the SPH estimate of the conductivity becomes ever more accurate.

\section{Spherical models for clusters of galaxies}

Observed temperature profiles of clusters of galaxies are often
characterised by a central decline of temperature, while otherwise
appearing fairly isothermal over the measurable radial range. Provided the
conductivity is non-negligible, there should hence be a conductive heat
flux into the inner parts, which would then counteract central cooling
losses.  Motivated by this observation, \cite{Zak03} (ZN henceforth) have
constructed simple analytic models for the structure of clusters, invoking
as a key assumption a local equilibrium between conduction and cooling.
Combined with the assumption of hydrostatic equilibrium and spherical
symmetry, they were able to quite successfully reproduce the temperature
profiles of a number of observed clusters.

We here use the model of ZN as a test-bed to check the validity of our
conduction modelling in a realistic cosmological situation. In addition, the
question for how long the ZN solution can be maintained is of immediate
interest, and we address this with our simulations of spherical clusters as
well.

For definiteness, we briefly summarise the method of ZN for constructing a
cluster equilibrium model. The cluster is assumed to be spherically
symmetric, with the gas of density $\rho(r)$ and pressure $P(r)$ being in
hydrostatic equilibrium, i.e.
\begin{equation}
  \frac{1}{\rho}\frac{\unit{d}P}{\unit{d}r} = -
  \frac{\unit{d}\Phi}{\unit{d}r},
\label{eqn:pressure}
\end{equation}
where the gravitational potential $\Phi$ is given by
\begin{equation}
  \frac{1}{r^2}\frac{\unit{d}}{\unit{d}r} \left( r^2
    \frac{\unit{d}\Phi}{\unit{d}r} \right) = 4 \pi G \left( \rho_{\rm
      DM} + \rho \right) \text{.}
\label{eqn:potential}
\end{equation}
The dark matter density $\rho_{\rm DM}(r)$ is described by a standard NFW
\citep*{NFW} halo, optionally modified by ZN with the introduction of a
small softened dark matter core. The structure of the dark halo can then be
fully specified by the virial radius $r_{\rm 200}$, defined as the radius
enclosing a mean overdensity of 200 with respect to the critical density,
and a scale length $r_s$. 

The radial heat flux $F$ due to electron conduction is given by
\begin{equation}
  F = - \kappa \frac{\unit{d}T}{\unit{d}r},
  \label{eqn:ZN_flow}
\end{equation}
where the conductivity $\kappa$ is taken to be a constant fraction of the
Spitzer value.  In order to balance local cooling losses, we require
\begin{equation}
  \frac{1}{r^2} \frac{\unit{d}}{\unit{d}r}\left( r^2 F \right) = - j,
\label{eqn:rates}
\end{equation}
where $j$ is describing the local gas cooling rate. If cooling is dominated
by thermal bremsstrahlung, $j$ can be approximated as
\begin{equation}
  j = 2.1 \times 10^{-27} n_e^2\, T^{1/2}\, \unit{ergs cm}^{-3}
  \unit{s}^{-1}.
  \label{eqn:ZN_cool}
\end{equation}
The electron number density $n_e$ herein correlates with the gas
density $\rho$ and depends on the hydrogen mass fraction $X$ we assume
for the primordial matter. With a proton mass of $m_p$, it equals
  $n_e = \rho ( X + 1 ) / ( 2 m_p)$.
Following ZN, we use this simplification in setting up our cluster models.

Equations (\ref{eqn:pressure}) to (\ref{eqn:ZN_cool}) form a system of
differential equations that can be integrated from inside out once
appropriate boundary conditions are specified.  We adopt the values for
central gas density and temperature, total dark matter mass, and scale
radius determined by ZN for the cluster A2390 such that the resulting
temperature profile matches the observed one well in the range $0\le r \le
R_{\rm out}$, where $R_{\rm out}$ was taken to be $2\, r_s$.

\begin{figure}
\includegraphics[width=8.5cm]{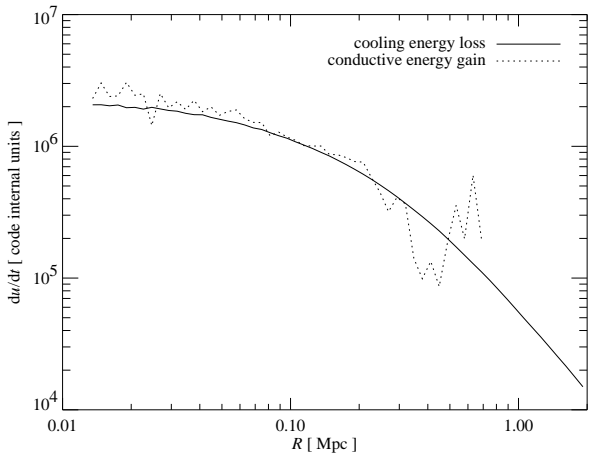}
\caption{Local cooling (solid line) and conductive heating rates
  (dotted line) as a function of radius in our model for the cluster
  A2390, after a time of $t=0.15\,\Gyr$.  It is seen that both
  energy transfer rates cancel each other with reasonable
  accuracy. \label{fig:cc-0.1}}
\end{figure}

Integrating the system of equations out to $R_{\rm out}$, we obtain a
result that very well matches that reported by ZN.  However, we also
need to specify the structure of the cluster in its outer parts in
order to be able to simulate it as an isolated system. ZN simply
assumed this part to be isothermal at the temperature $T_{\rm
out}=T(R_{\rm out})$, thereby implicitly assuming that the equilibrium
condition between cooling and conduction invoked for the inner parts
is not valid any more. It is a bit unclear why such a sudden
transition should occur, but the alternative assumption, that equation
(\ref{eqn:rates}) holds out to the virial radius, clearly leads to an
unrealistic global temperature profile.  In this case, the temperature
would have to monotonically increase out to the outermost
radius. Also, since the cumulative radiative losses out to a radius
$r$ have to be balanced by a conductive heat flux of equal magnitude
at that radius, the heat flux also monotonically increases, such that
all the energy radiated by the cluster would have to be supplied to
the cluster at the virial radius.

We nevertheless examined both approaches, i.e.~we construct cluster
models where we solve equations
(\ref{eqn:pressure})-(\ref{eqn:ZN_cool}) for the whole cluster out to
a radius of $r_{\rm 200}$, and secondly, we construct initial
conditions where we follow ZN by truncating the solution at $2\,
r_{\rm s}$, continuing into the outer parts with an isothermal
solution that is obtained by dropping equations (\ref{eqn:ZN_flow})
and (\ref{eqn:rates}).  Since conduction may still be important at
radii somewhat larger than $2\, r_s$, these two models may hence be
viewed as bracketing the expected real behaviour of the cluster in a
model where cooling is balanced by conduction.

\begin{figure*}
  \includegraphics[width=8.0cm]{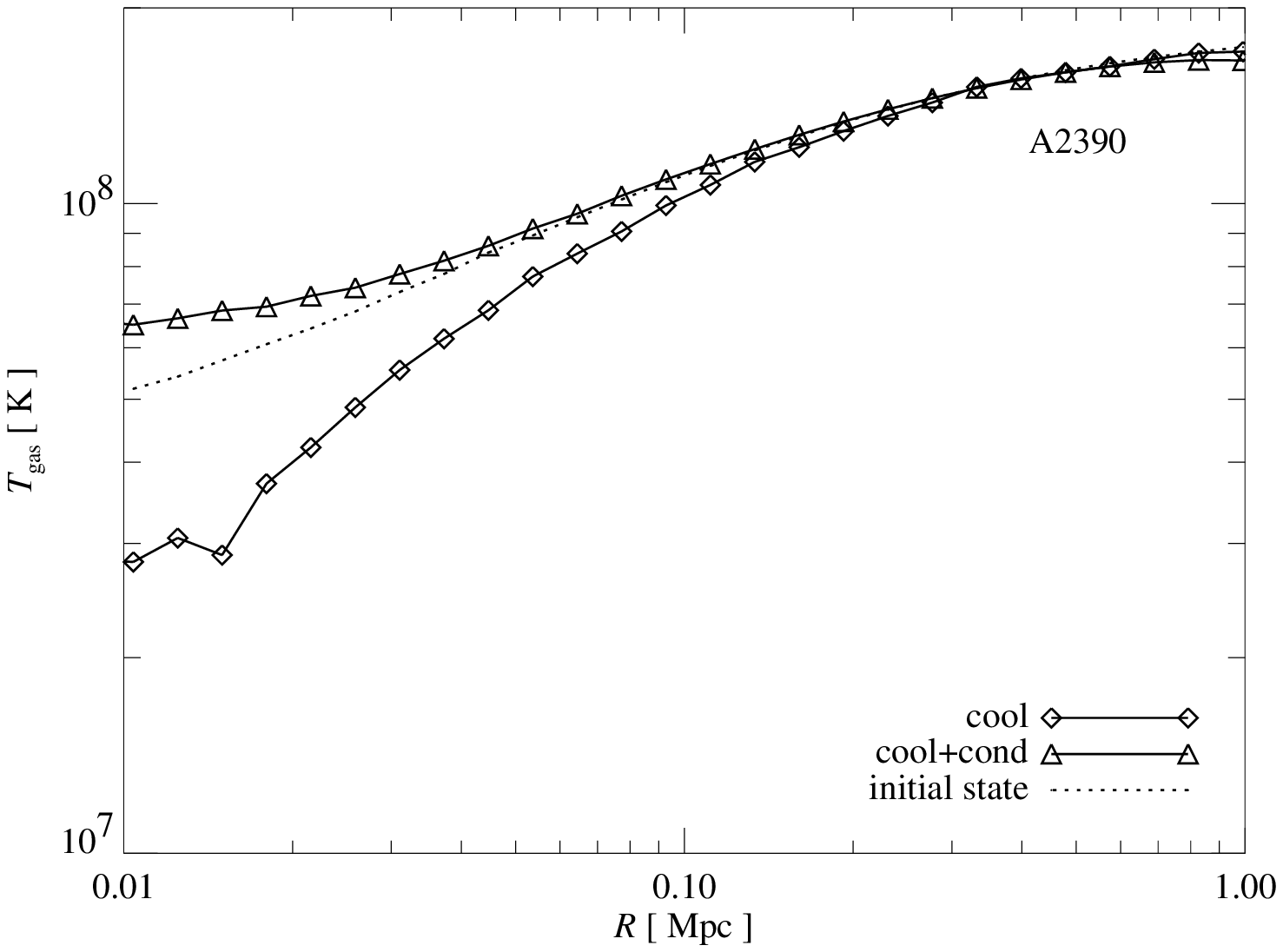}
  \includegraphics[width=8.0cm]{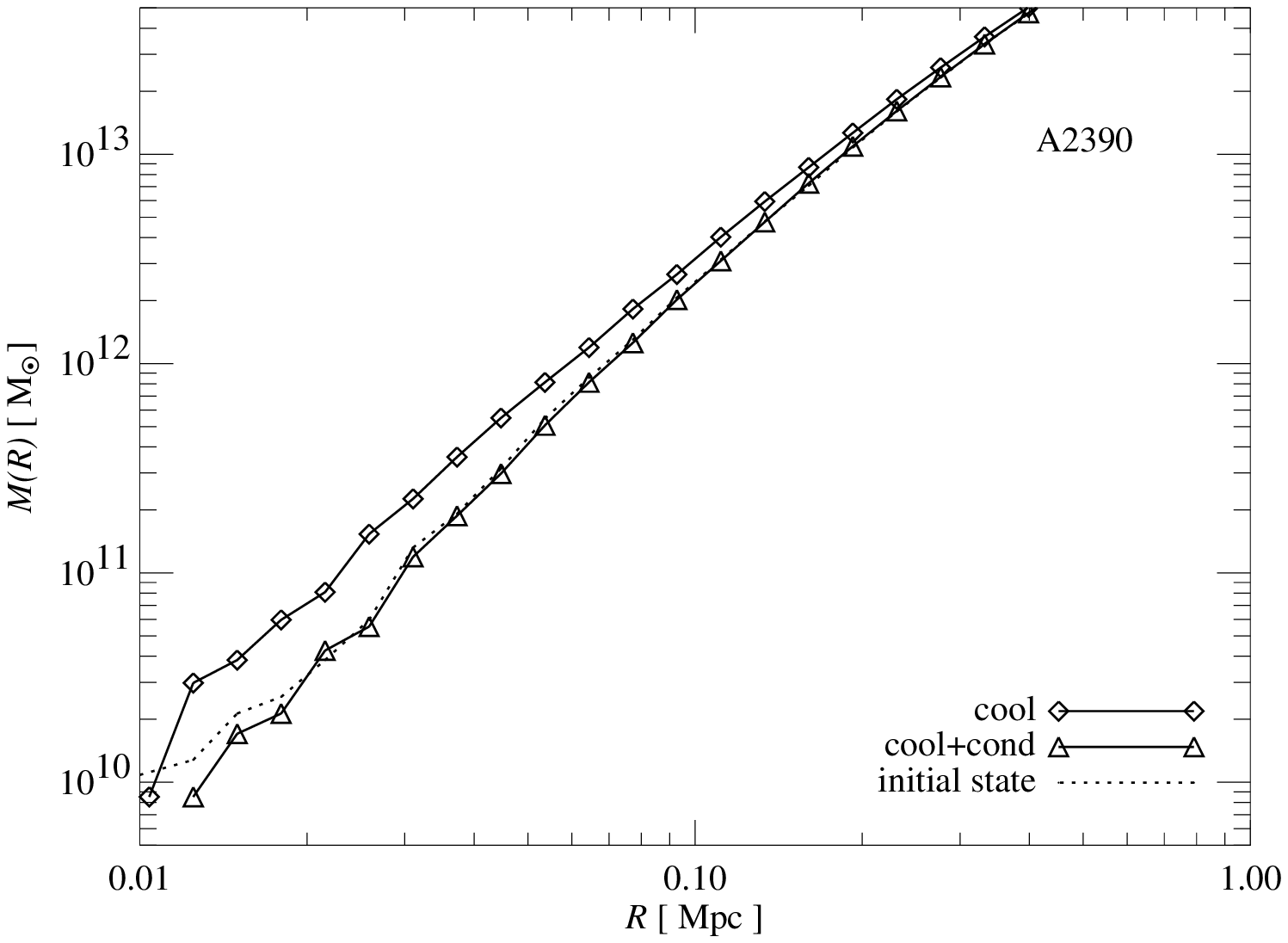}
  \caption{Temperature profile (left) and cumulative mass profile
    (right) of our model for the cluster A2390, after a simulated time
    of $0.6\,\Gyr$. We compare simulations with (triangles) and
    without (diamonds) thermal conduction using a conductivity of
    $0.3\,\kappa_{\rm sp}$.  It is seen that without
    conduction, the inner regions of the cluster cool down
    significantly, causing mass to sink towards the centre as central
    pressure support is partially lost. If thermal conduction is
    included, the central cooling losses are offset by conductive
    heating from outer regions, preventing any significant change of the
    baryonic mass profile. In fact, in this case we even observe a slow
    secular evolution towards higher temperature in the inner
    parts.\label{fig:A2390}}
\end{figure*}

For the plasma conductivity, we assumed a value of $0.3\,\kappa_{\rm sp}$,
as proposed by ZN in their best-fit solution for A2390. Note that this
value implies that a certain degree of suppression of conduction by
magnetic fields is present, but that this effect is (perhaps
optimistically) weak, corresponding to what is expected for
chaotically tangled magnetic fields \citep{Narayan01}.

Having obtained a solution for the static cluster model, we realized
it as 3D initial conditions for {\small GADGET}. We used $2\times
10^5$ gas particles with a total baryonic mass of around $8.6 \times
10^{14}\ \unit{M}_{\odot}$.
For simplicity,
we described the NFW dark matter halo of mass $2.4\times 10^{15}\,{\rm M_{\odot}}$ as a static
potential.  We then simulated the evolution of the gas subject to
self-gravity (with a gravitational softening length of
$3\,\unit{kpc}$), radiative cooling, and thermal conduction, but
without allowing for star formation and associated feedback processes.

By construction, we expect that thermal conduction will be able to
offset radiative cooling for these initial conditions, at least for
some time.  This can be verified in Figure~\ref{fig:cc-0.1}, where we
plot the local cooling rate and conductive heating rate as a function
of radius.  Indeed, at time $0.15\,{\rm Gyr}$, after a brief
initial relaxation period, the two energy transfer rates exhibit the
same magnitude and cancel each other with good accuracy. This
represents a nice validation of our numerical implementation of
thermal conduction with the temperature-dependent Spitzer-rate.

As the simulation continues, we see that the core temperature of the
cluster slowly drifts to somewhat higher temperature. A secular
evolution of some kind is probably unavoidable, since the balance
between cooling and conductive heating cannot be perfectly static. The
cluster of course still loses all the energy it radiates, which
eventually must give rise to a slow quasi-static inflow of gas, and a
corresponding change of the inner structure of the cluster.  Because
of the different temperature dependences of cooling and conductive
heating, it is also not clear that the balance between cooling
and conductive heating represents a stable dynamical state.
While a stability analysis by ZN and \citet{Kim2003} suggests that
thermal instability is sufficiently suppressed in models with
conduction, \citet{Soker2003} argues that local perturbations in the
cooling flow region will grow to the non-linear regime rather quickly
and that a steady solution with a constant heat conduction may
therefore not exist.

In any case, it is clear that the inclusion of conduction strongly
reduces central mass drop out due to cooling. This is demonstrated
explicitly in Figure~\ref{fig:A2390}, where we show the temperature
and cumulative baryonic mass profiles of the cluster after a time of
$0.6\,{\rm Gyr}$. We compare these profiles also to an identical
simulation where conduction was not included. Unlike the conduction
run, this cooling-only simulation shows a very substantial
modification of its temperature profile in the inner parts, where the
core region inside $\sim 0.1\,{\rm Mpc}$ becomes much colder
than the initial state.  We can also see that the cooling-only
simulation has seen substantial baryonic inflow as a result of central
mass drop-out.  Within $0.6\,{\rm Gyr}$, the baryonic mass
enclosed in a sphere with radius $50\,{\rm kpc}$ has increased
by 86 percent in this simulation, clearly forming a cooling flow.
This contrasts strongly with the run that includes conduction,
where beyond a cental distance of $\sim 0.02\,{\rm Mpc}$, there
is no significant difference in the cumulative mass profile between $t
= 0$ and $t = 0.6\,{\rm Gyr}$.

We also checked that runs with a reduced conductivity show a likewise
reduced effect, suppressing the core cooling and matter inflow to a
lesser extent than with our default model with $0.3\,\kappa_{\rm sp}$.

Finally, we note that the results discussed here are quite insensitive to
whether we set-up the initial conditions with an isothermal outer part, or
whether we continue the equilibrium solution to the virial
radius. There is only a small difference in the long term evolution of
the solution.

The model presented here helps us to verify the robustness of
our conduction implementation when a temperature dependent
conductivity is used, but it does not account for the time dependence
of conduction expected in the cosmological context.
The static, spherically symmetric construction
does not reflect the hierarchical growth of clusters that is a central
element of currently favoured cosmologies. To be able to make reliable
statements about the effects of thermal conduction on galaxy clusters,
one therefore has to trace their evolution from high redshift
to the present, which is best done in full cosmological simulations.
Here, the hierarchical growth of clusters implies that the
conductivity is becoming low early on, when the ICM temperature is
low, and becomes only large at late times when the cluster forms.

\section{Cosmological Cluster Simulations}

\begin{figure*}
  \includegraphics[width=5cm]{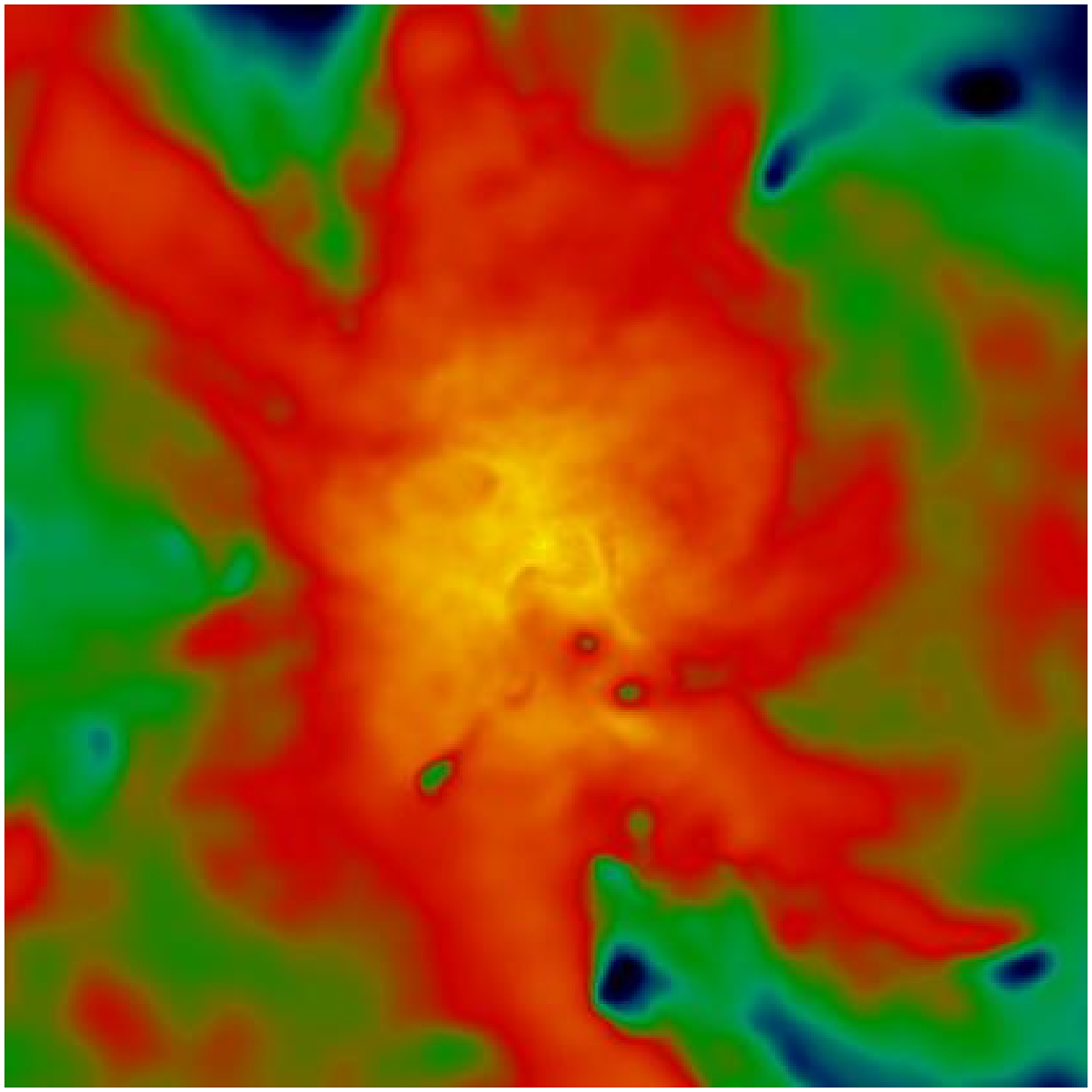}
  \includegraphics[width=5cm]{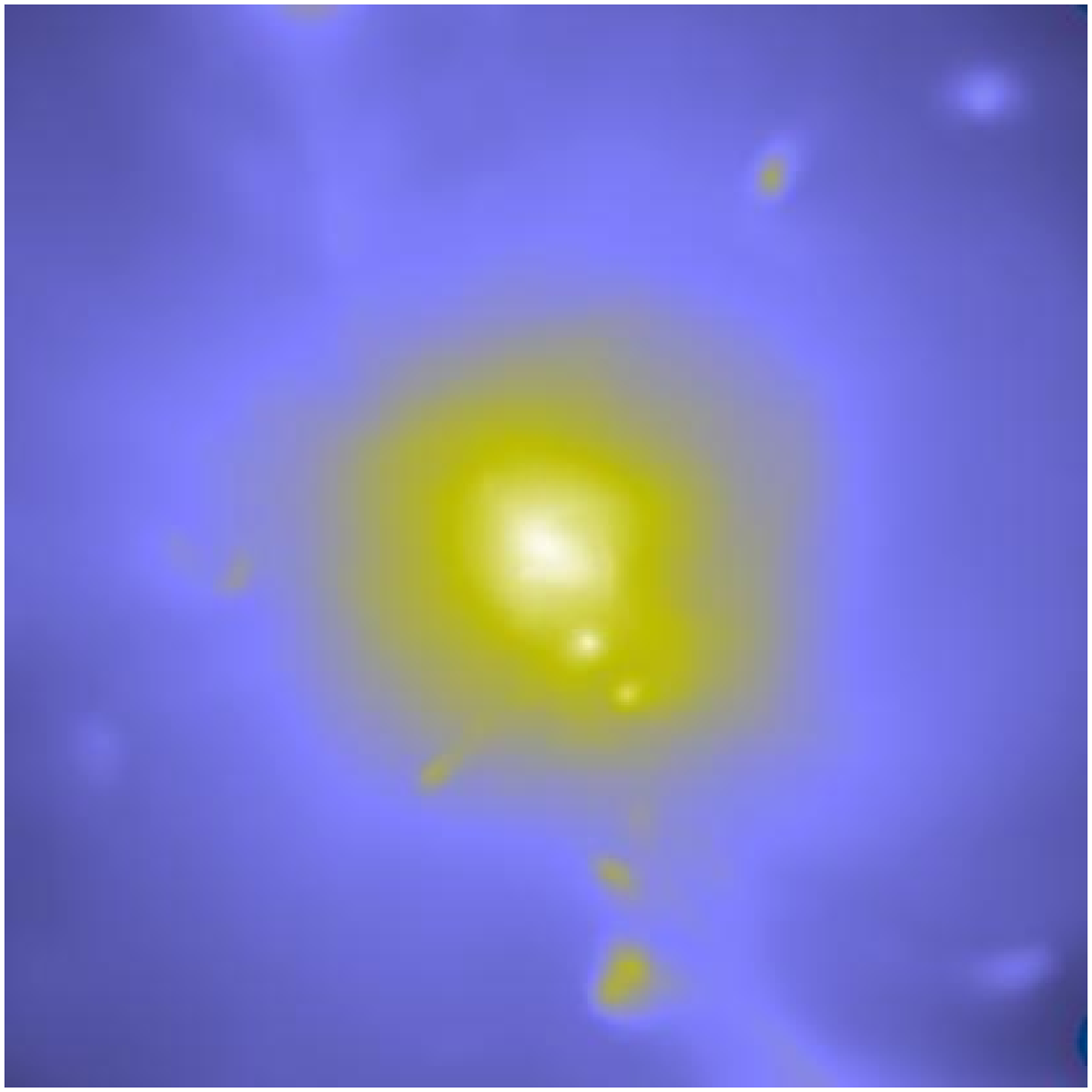}
  \includegraphics[width=5cm]{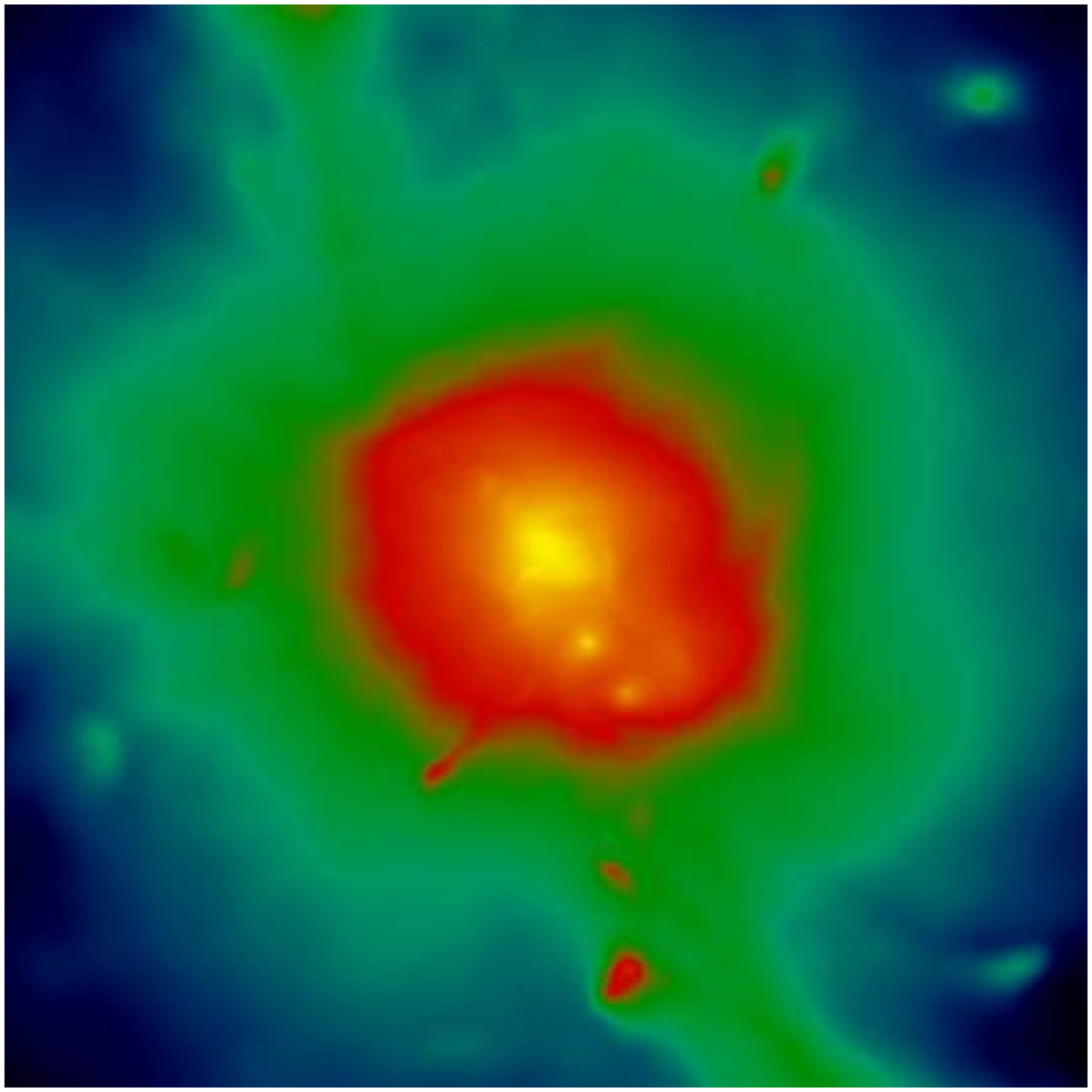}
  \includegraphics[width=5cm]{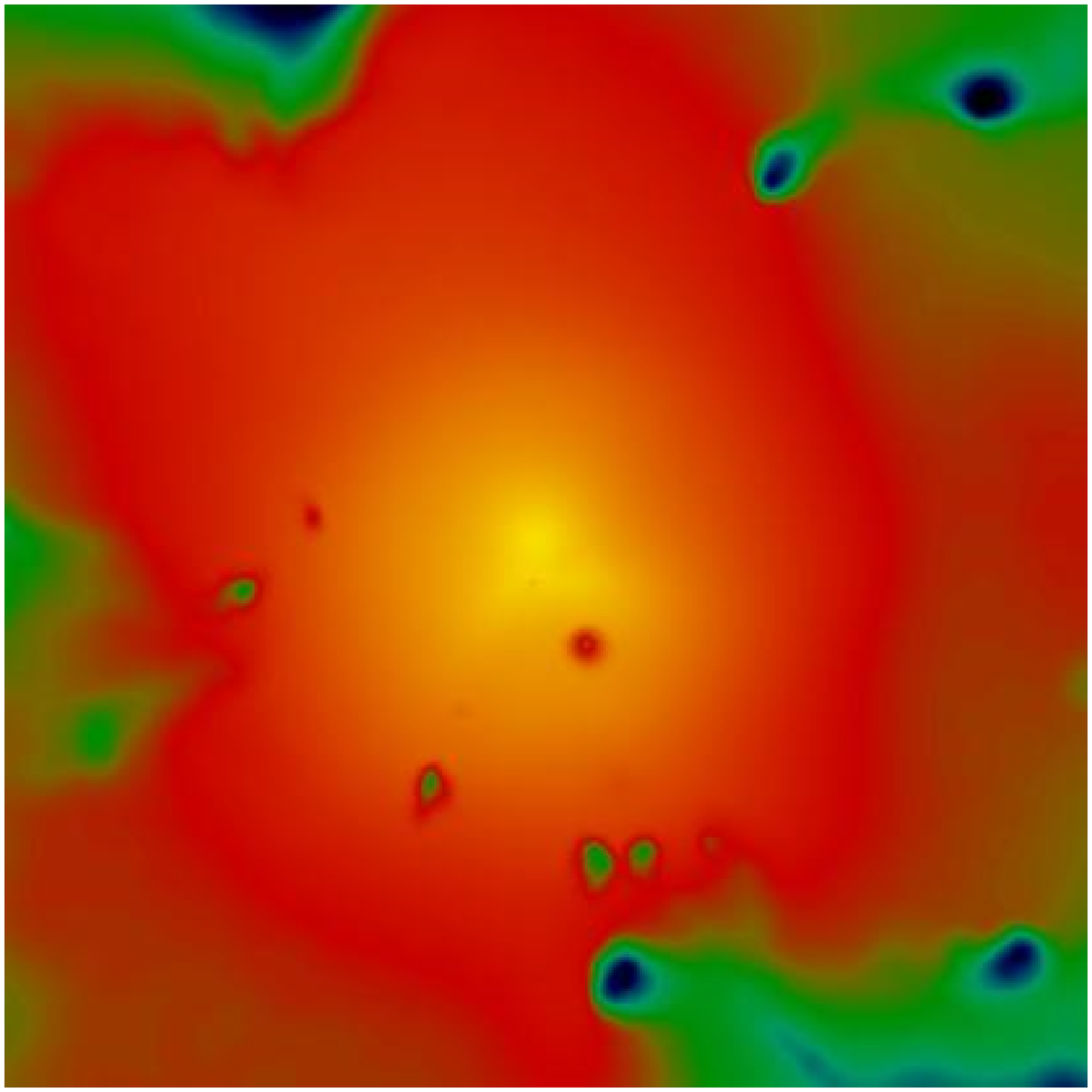}
  \includegraphics[width=5cm]{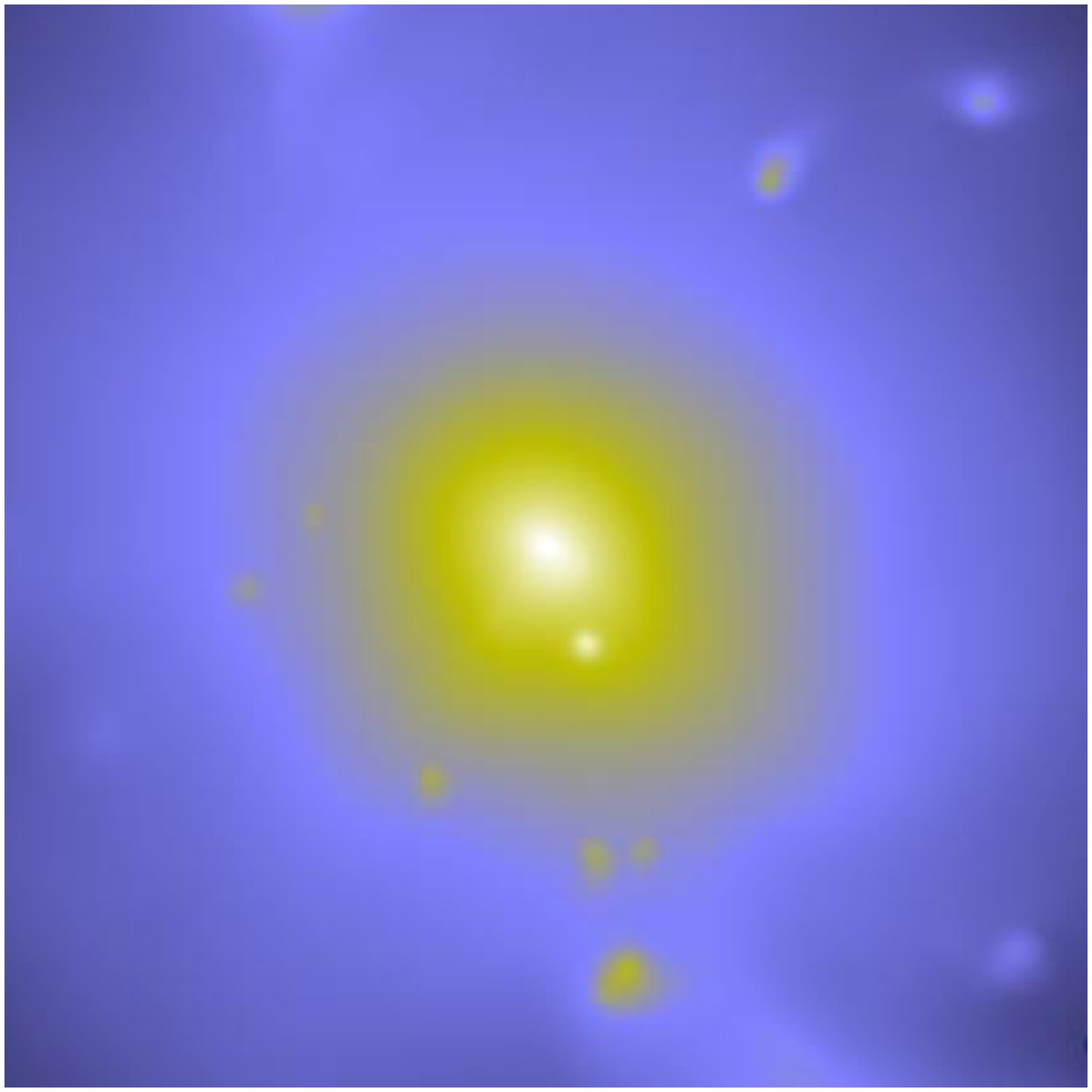}
  \includegraphics[width=5cm]{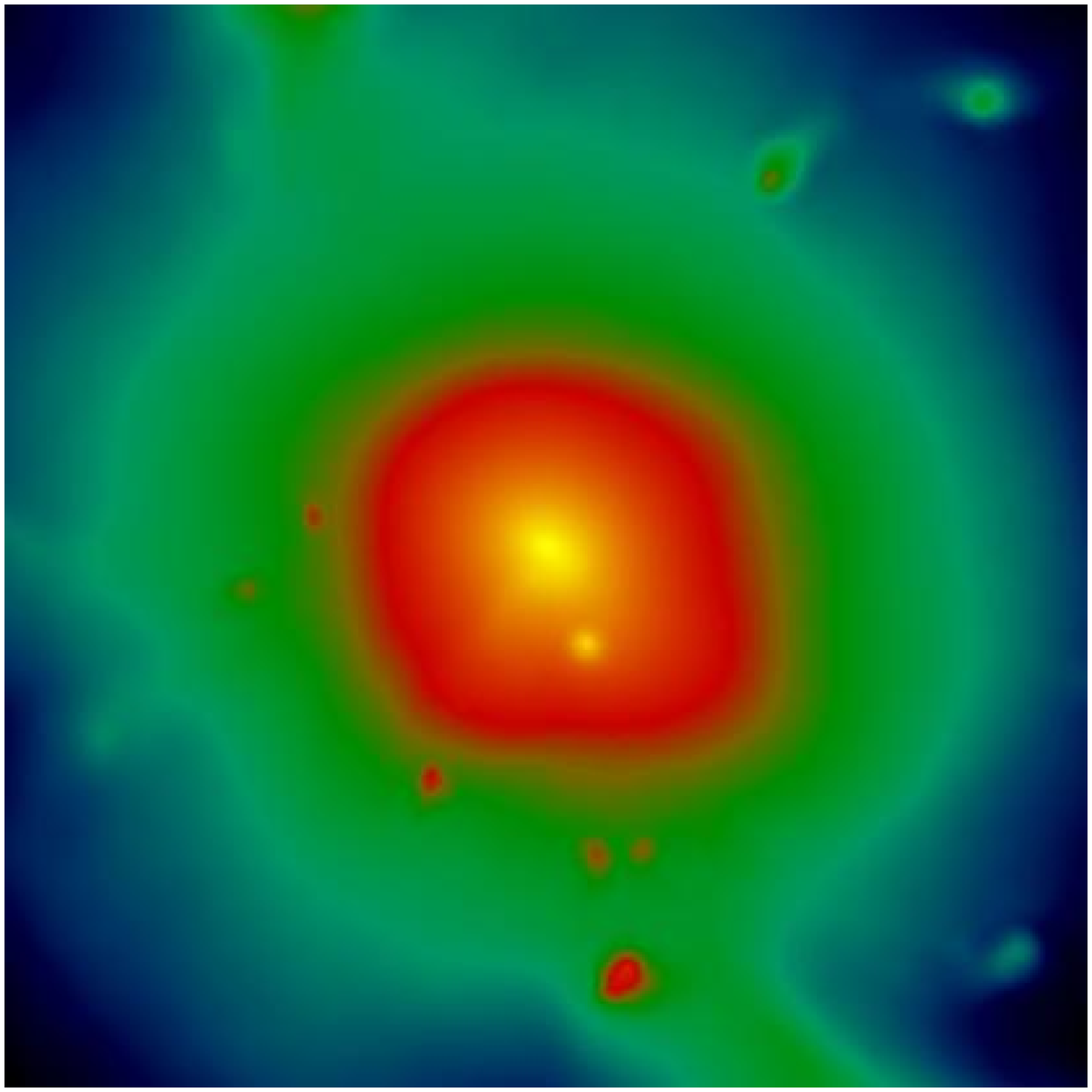}
  \includegraphics[width=5cm]{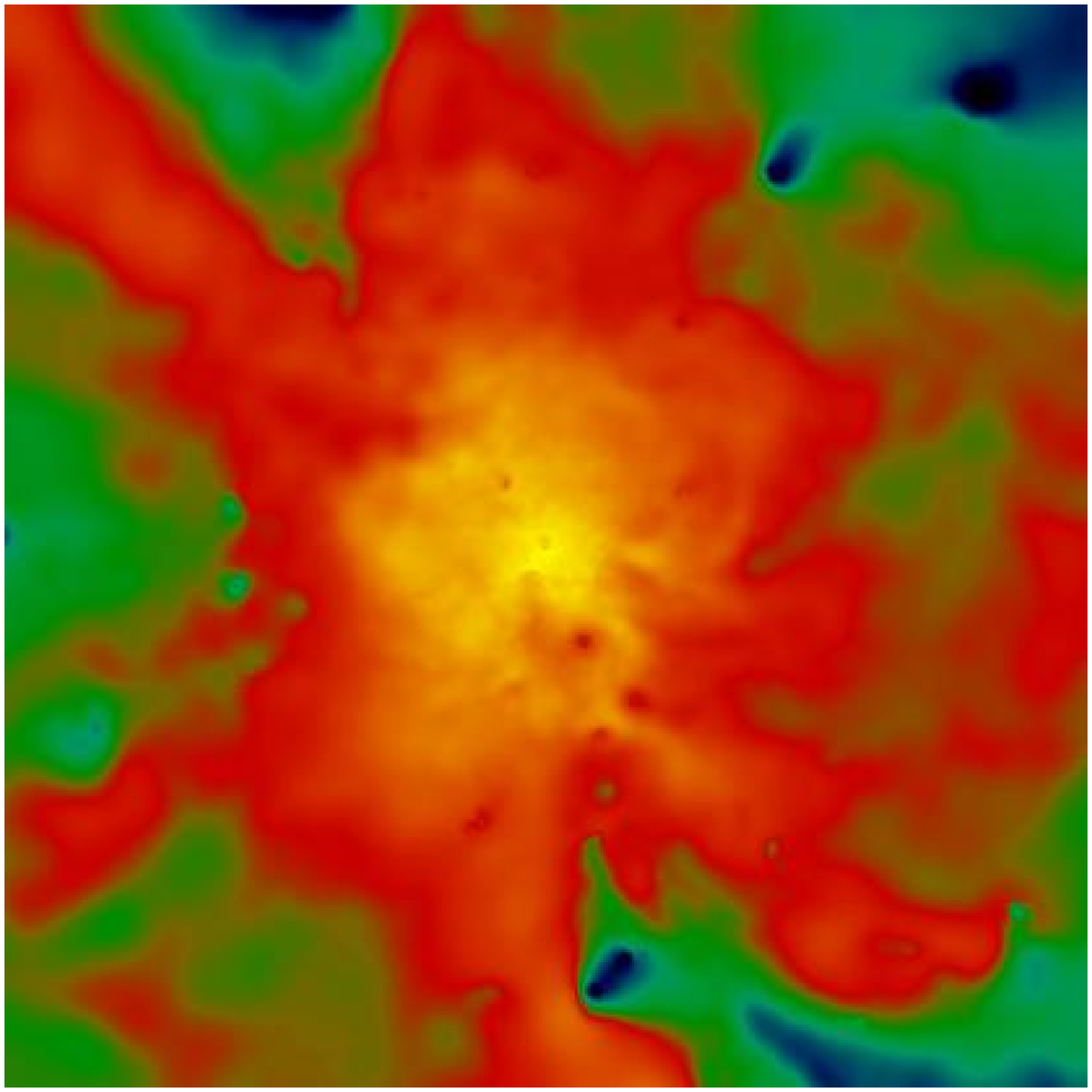}
  \includegraphics[width=5cm]{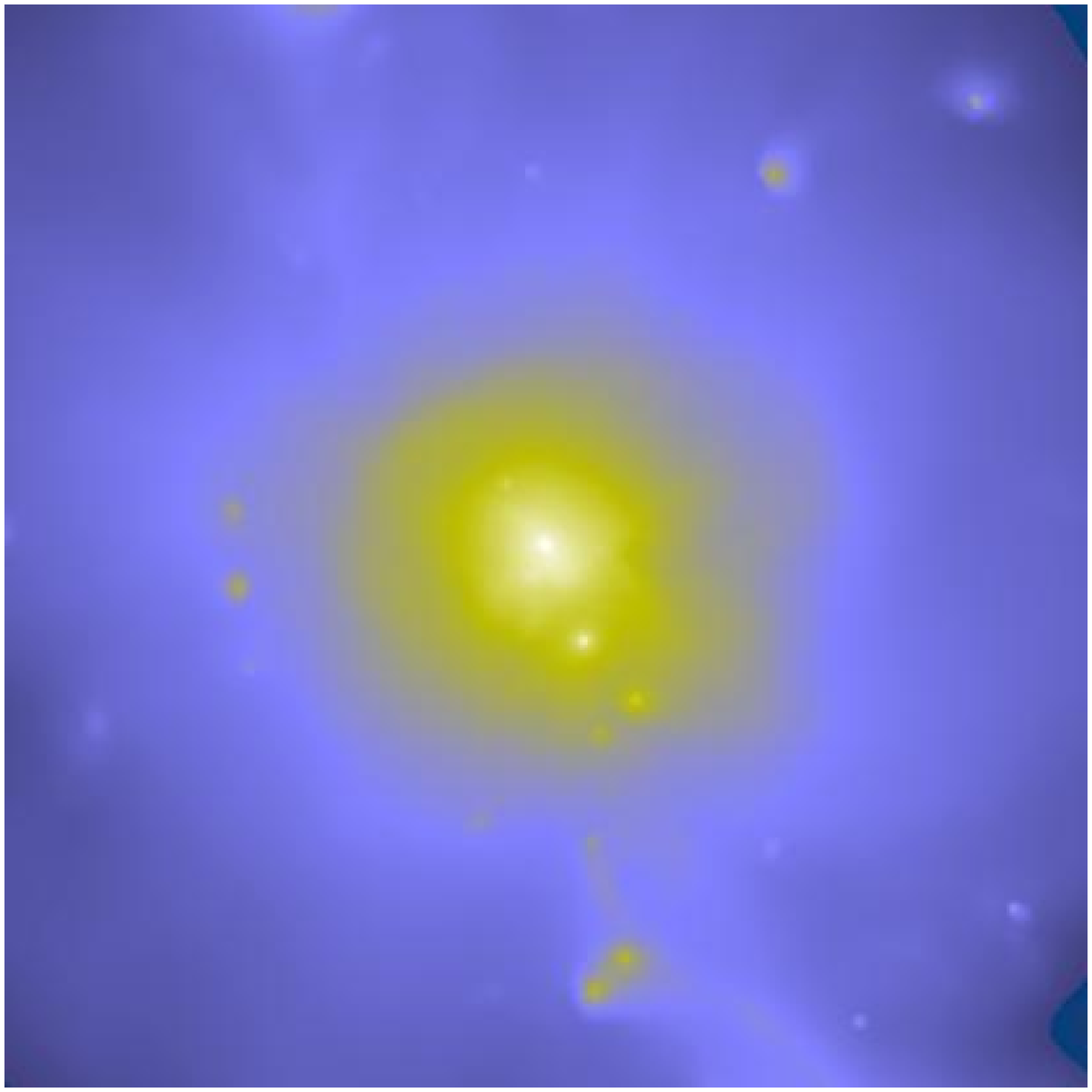}
  \includegraphics[width=5cm]{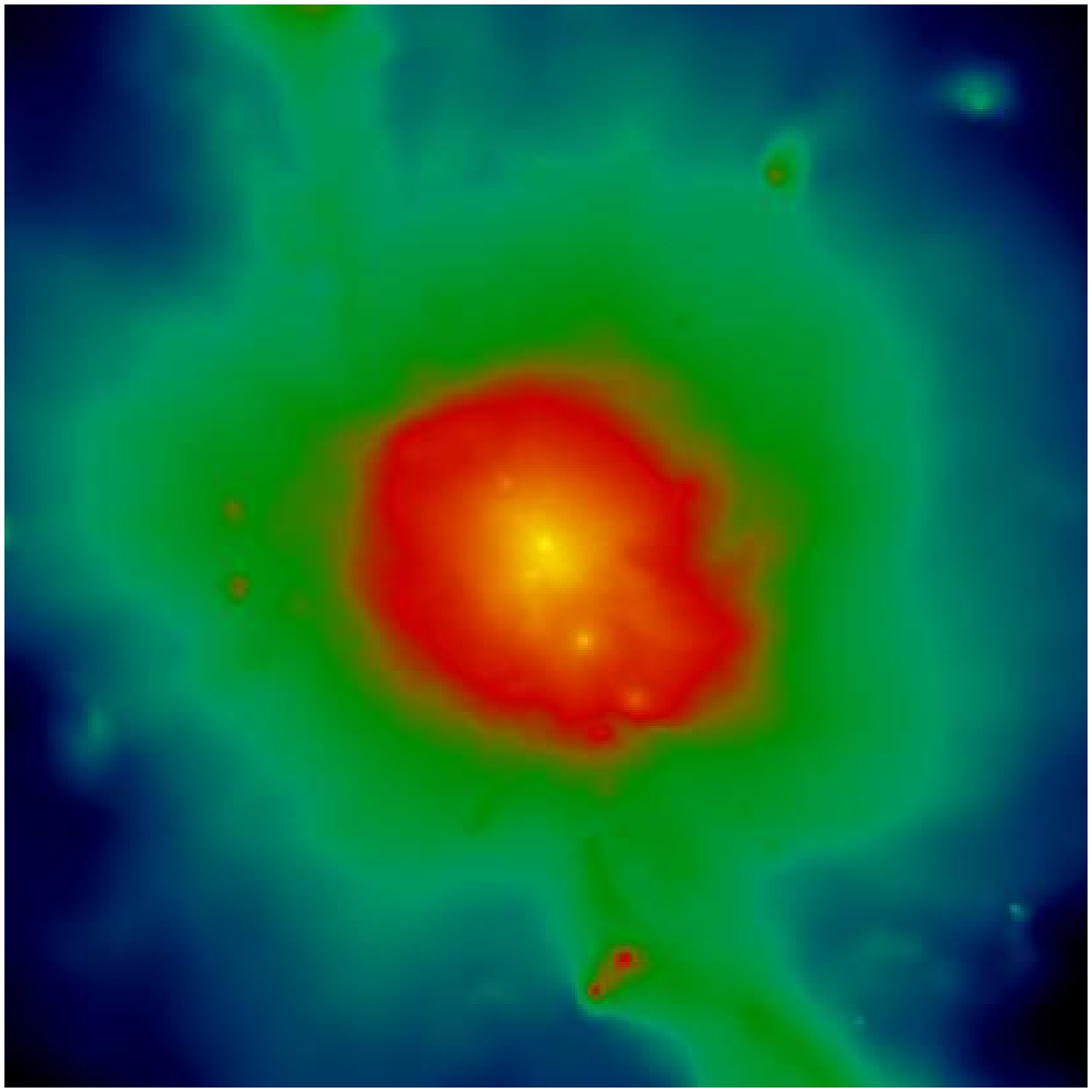}
  \includegraphics[width=5cm]{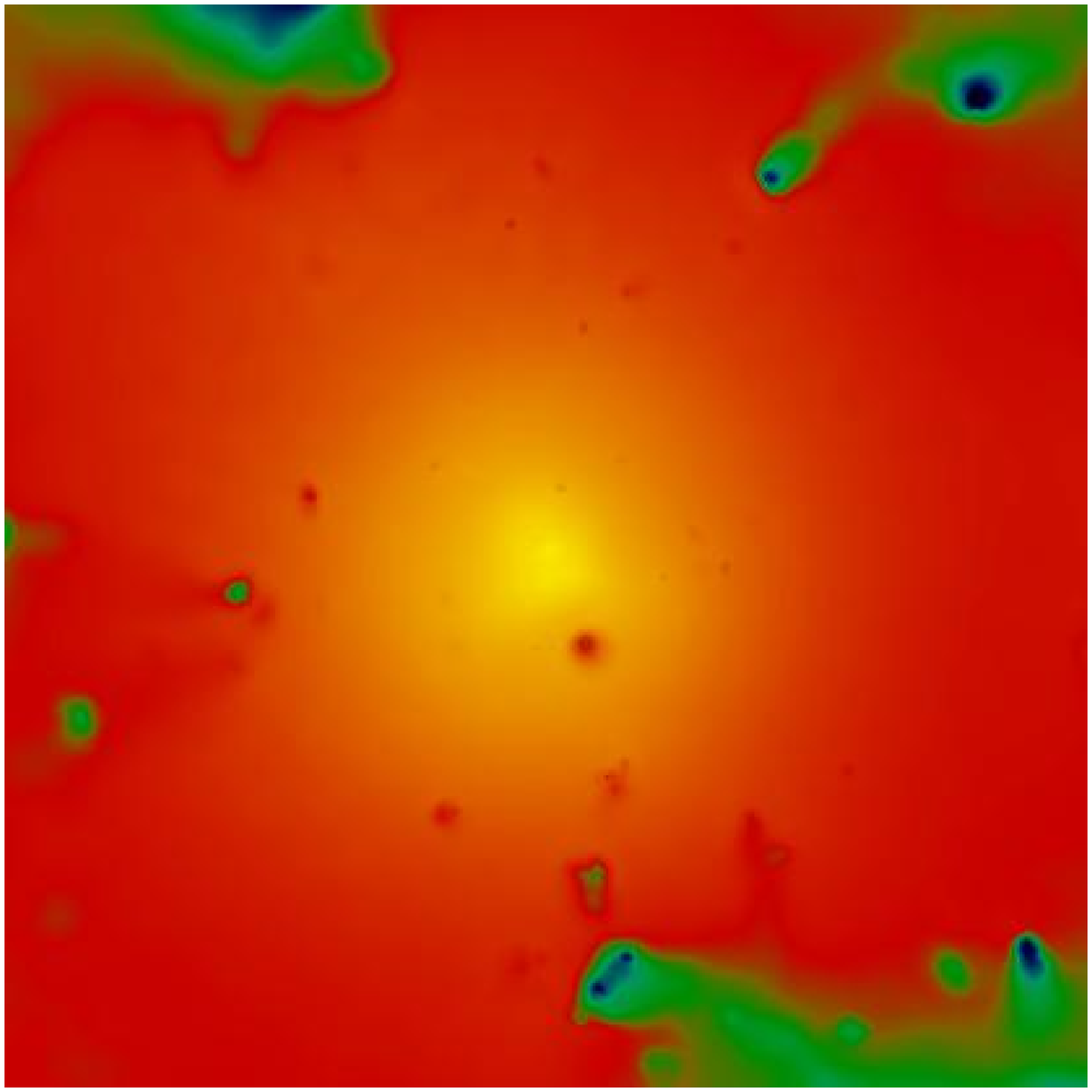}
  \includegraphics[width=5cm]{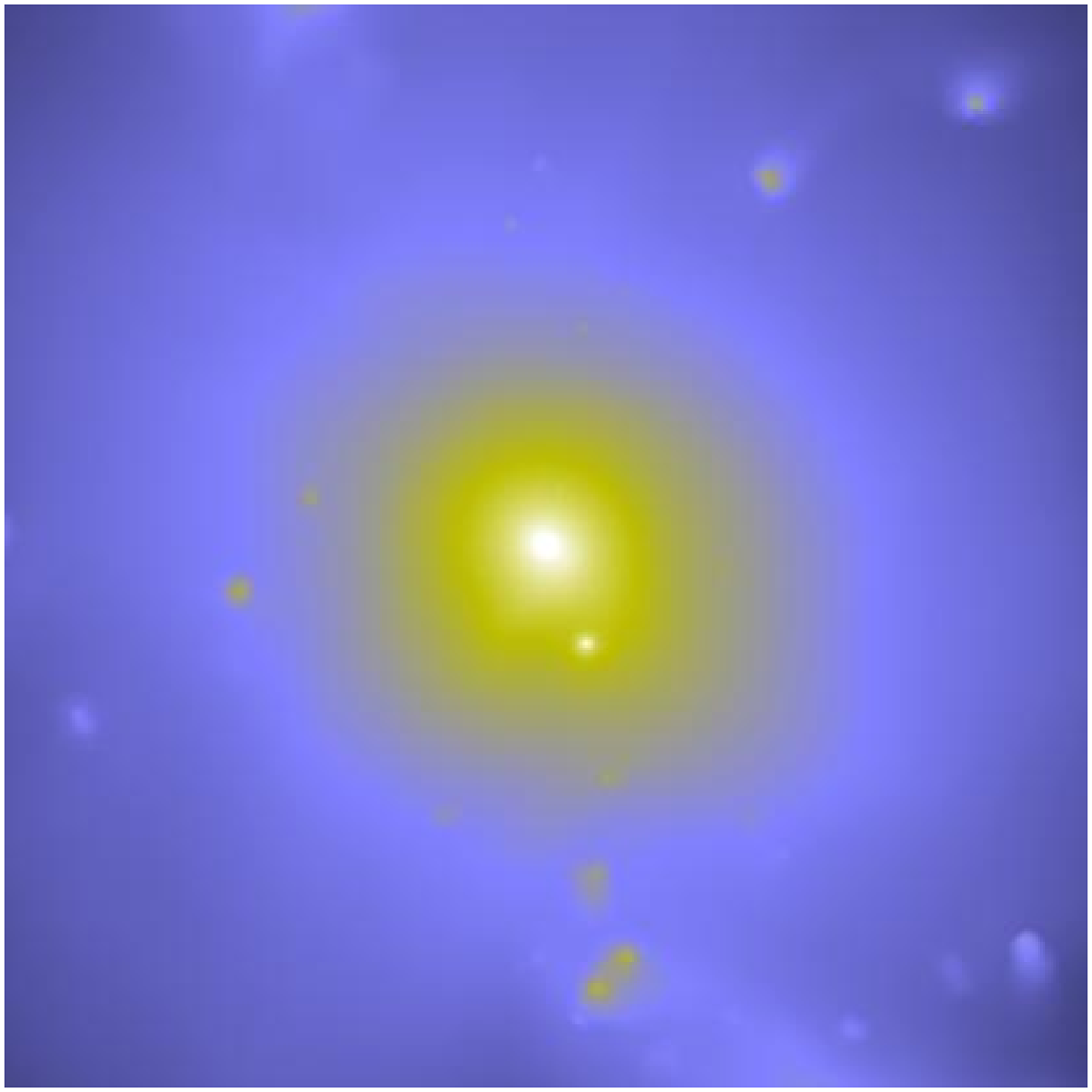}
  \includegraphics[width=5cm]{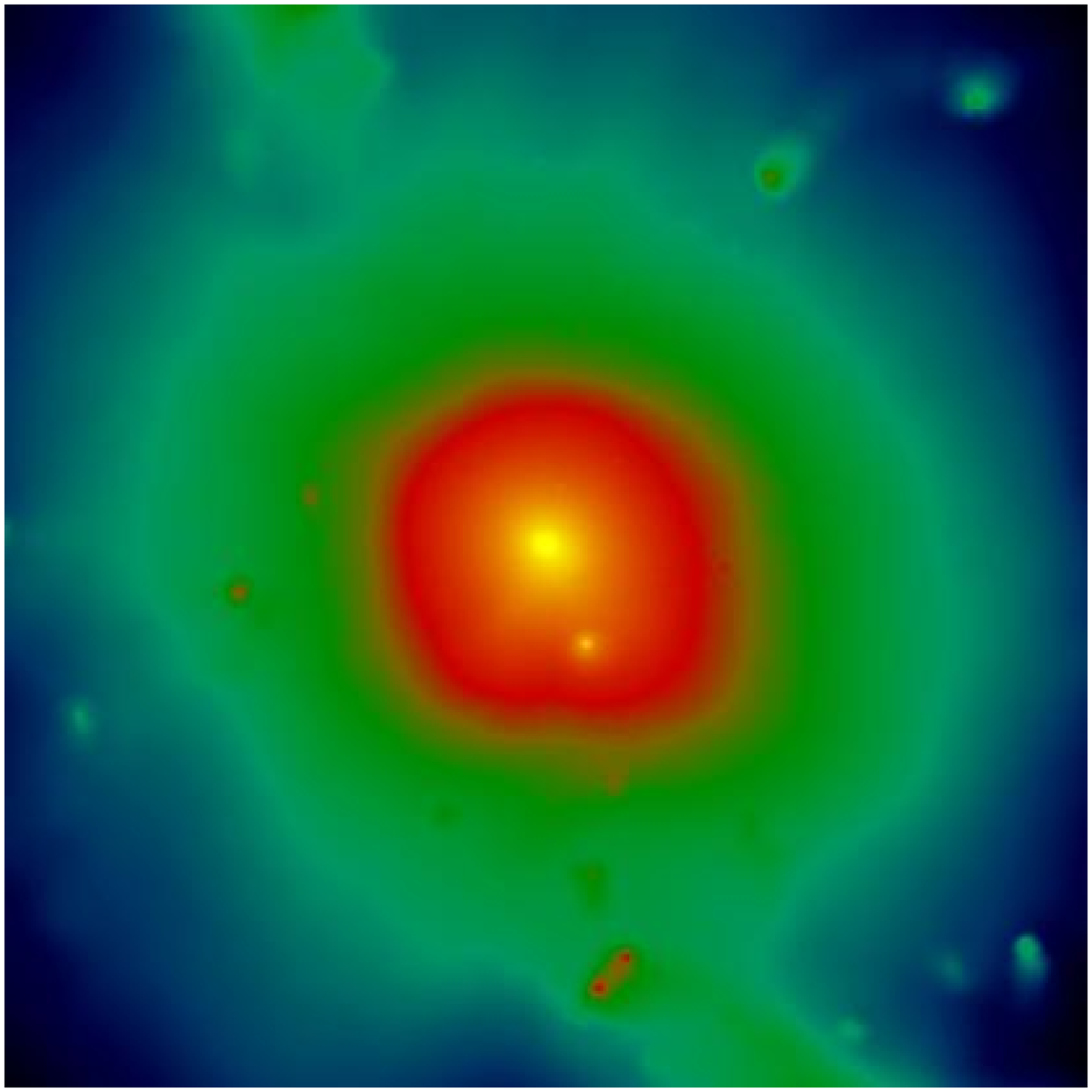}
  \includegraphics[width=5cm]{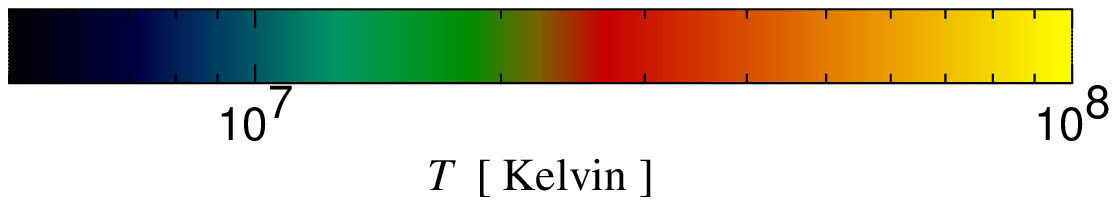}
  \includegraphics[width=5cm]{figures/empty_legend.eps}
  \includegraphics[width=5cm]{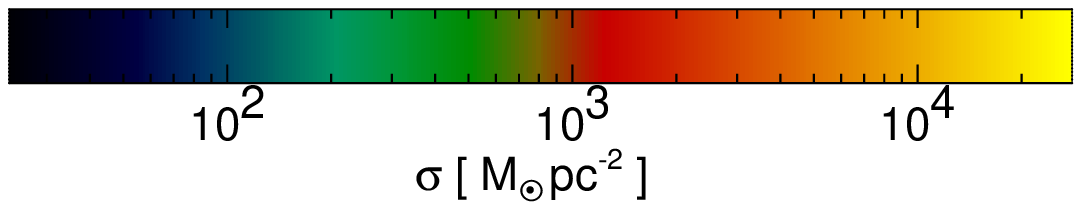}
  \caption{Projections of mass-weighted temperature (left column), X-ray
    emissivity (middle column), and gas mass density (right column) for
    our cluster simulations at $z=0.13$. From the top to the bottom row, we
    show the same cluster but simulated with different physics: Adiabatic
    gasdynamics only, adiabatic plus thermal conduction, radiative cooling
    and star formation, and finally, cooling, star formation and
    conduction.  Each panel displays the gas contained in a box of
    side-length $8.6\,{\rm Mpc}$ centred on the cluster. Full Spitzer
    conductivity was assumed.
    \label{figprojections}}
\end{figure*}

\begin{figure}
  \includegraphics[width=8.0cm]{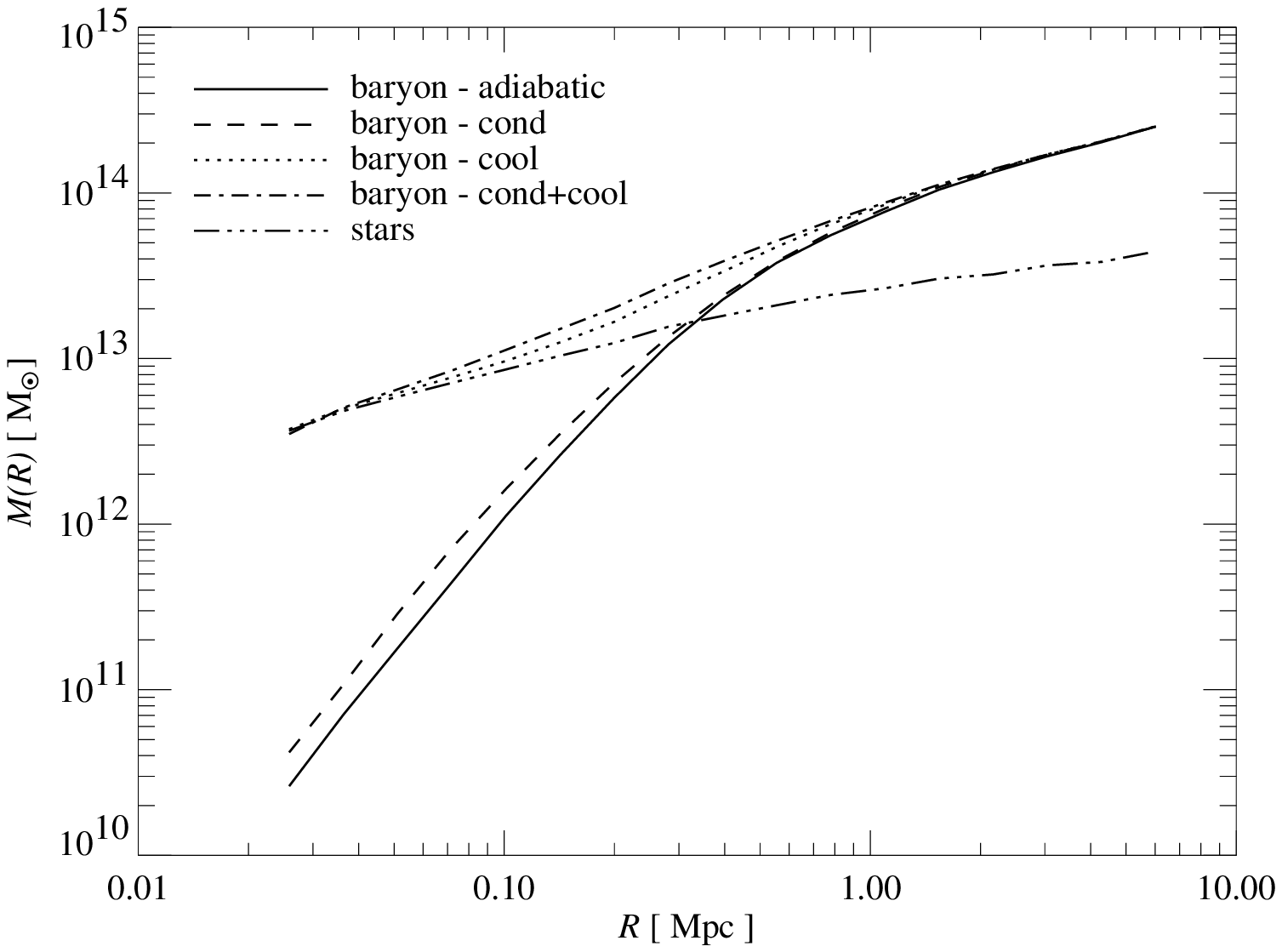}
  \includegraphics[width=8.0cm]{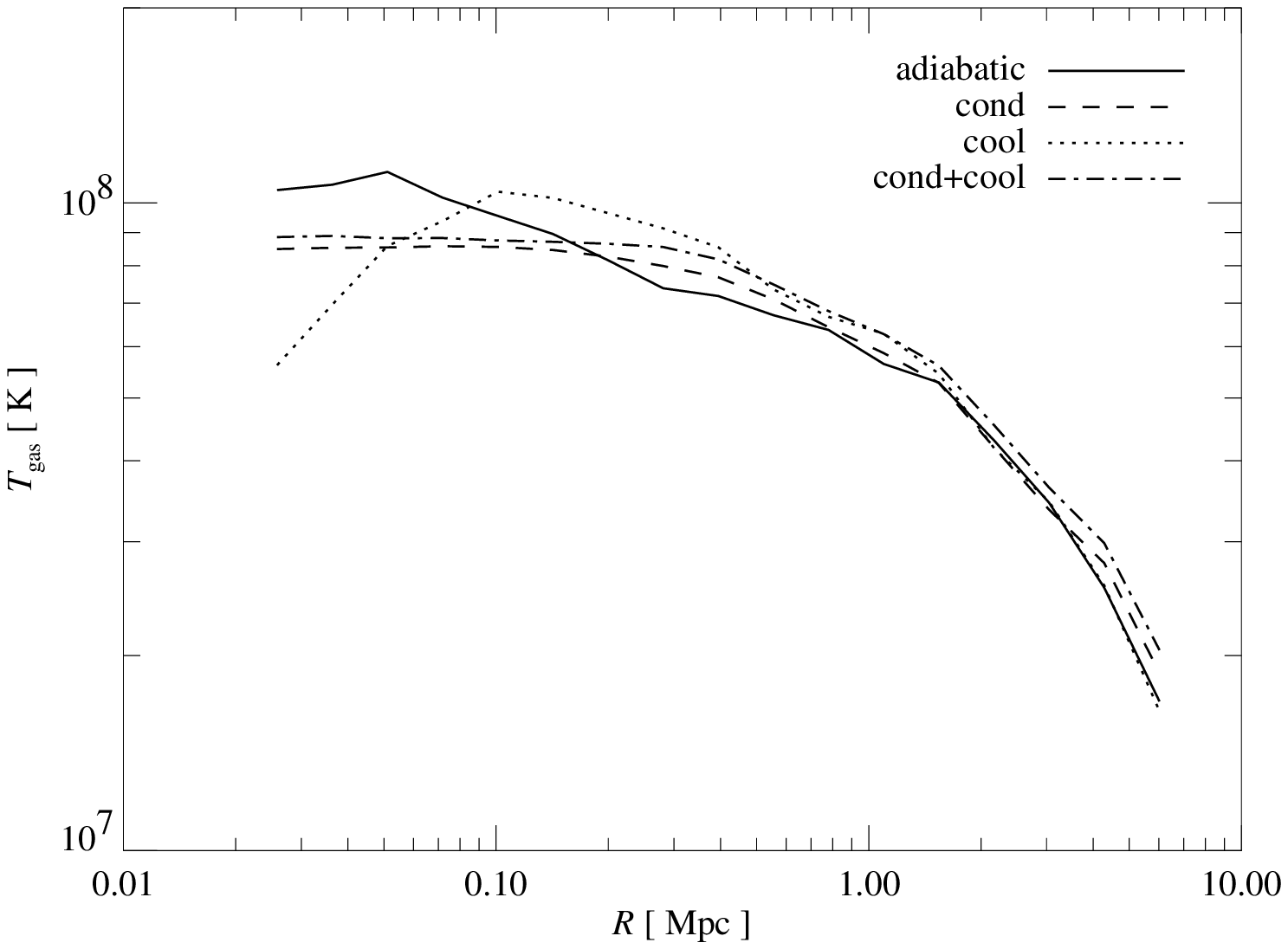}
  \includegraphics[width=8.0cm]{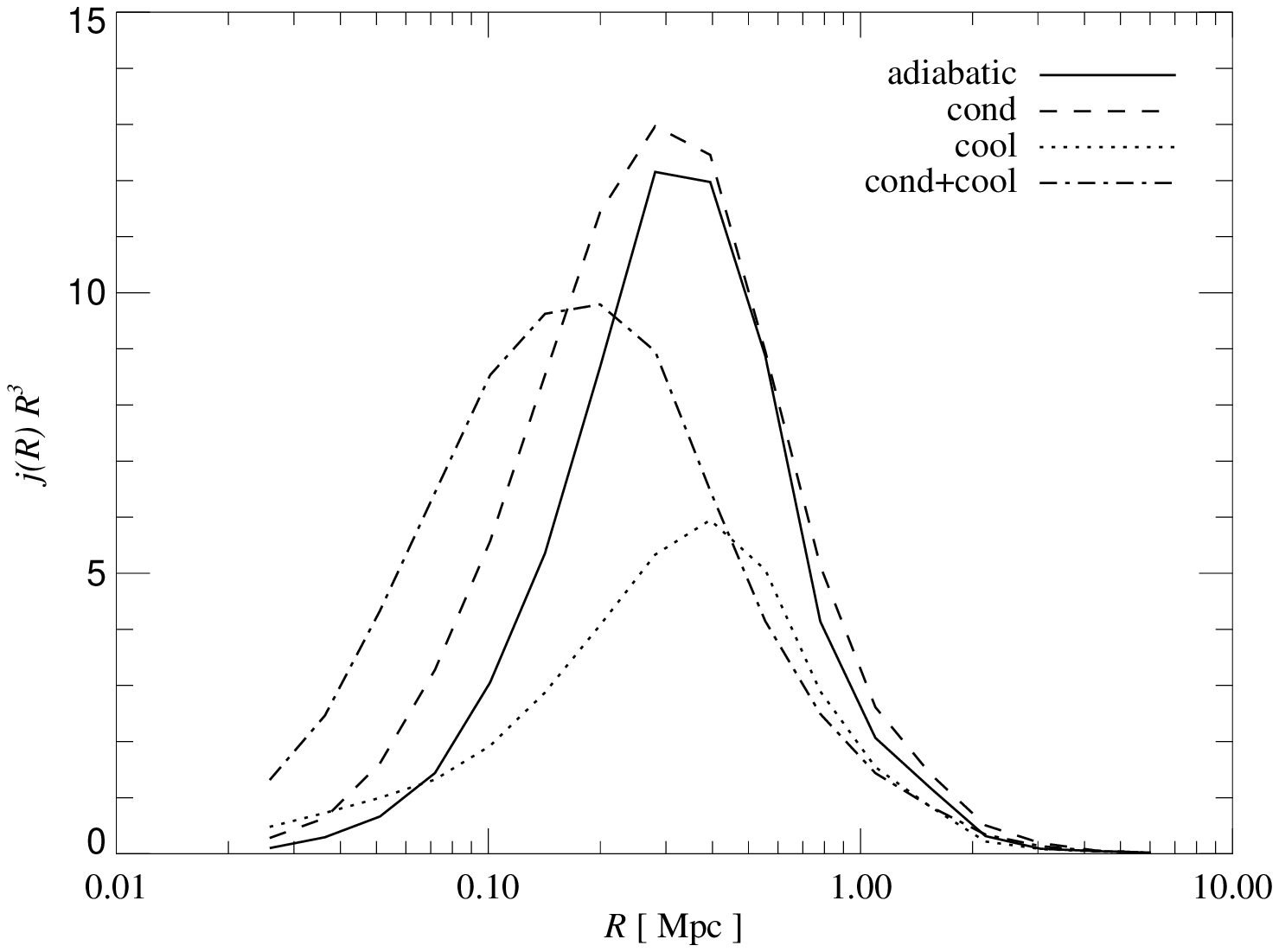}
  \caption{Cumulative baryonic mass profile (top), temperature profile
    (middle) and X-ray emissivity profile (bottom) of the
    simulated cluster at $z = 0.13$. In each panel, we compare the same
    cluster simulation run with different physical models for the gas:
    Adiabatic gasdynamics only, adiabatic plus thermal conduction,
    radiative cooling and star formation without conduction, and
    finally, cooling, star  formation and conduction. 
    For the models including conduction, full Spitzer conductivity was
    assumed. Note that the
    X-ray emissivity is plotted
    such that the area under the curve is proportional to the total
    bolometric X-ray luminosity.
    \label{fig:cosmocompare}}
\end{figure}

In this section, we apply our new numerical scheme for the treatment
of thermal conduction in fully self-consistent cosmological
simulations of cluster formation. We will in particular compare the
results of simulations with and without conduction, both for runs that
follow only adiabatic gas-dynamics, and runs that also include
radiative cooling of gas. While it is beyond the scope of this work to
present a comprehensive analysis of the effects of conduction in
cosmological simulations, we here want to investigate a set of small
fiducial runs in order to further validate the stability of our method
for real-world applications, and secondly, to give a first flavour of
the expected effects of conduction in simulated clusters.  A more
detailed analysis of cosmological implications of conduction is
presented in a companion paper \citep{Dolag04}.

We focus on a single cluster, extracted from the `GIF' simulation
\citep{Kau99} of the $\Lambda$CDM model, and resimulated using the
``zoomed initial conditions'' technique \citep{Tor97}. To this end,
the particles that make up the cluster at the present time are traced
back to their original coordinates in the initial conditions. The
Lagrangian region of the cluster identified in this way is then
resampled with particles of smaller mass, and additional small-scale
perturbations from the CDM power spectrum are added appropriately. Far
away from the cluster, the resolution is progressively degraded by
using particles of ever larger mass.

The cluster we selected has a virial mass of $1.1\times
10^{15}\,{\rm M}_\odot$. It is the same cluster considered in
the high-resolution study of \citet{Spr99b}, with our mass resolution
corresponding to their S1 simulation, except that we split each of the
$\sim 450000$ high resolution dark matter particles into a gas and a
dark matter particle (assuming $\Omega_b=0.04$), yielding a gas mass
resolution of $8.4 \times 10^9\,{\rm M_\odot}$.  The boundary
region was sampled with an additional 3 million dark matter
particles. We then evolved the cluster forward in time from a starting
redshift $z = 30$ to the present, $z = 0$, using a comoving
gravitational softening length of $20\,{\rm kpc}$.

We consider 4 different simulations of the cluster, each with
different physical models for the gas: (1) Adiabatic gasdynamics only,
(2) adiabatic gas and conduction, (3) radiative cooling and star
formation without conduction, and finally, (4) radiative cooling, star
formation and conduction. The simulations with cooling and star
formation use the sub-resolution model of \citet{SprHer03} for the
multi-phase structure of the ISM (without including the optional
feedback by galactic winds offered by the model).  In the two runs
with conduction, we adopt the full Spitzer rate for the
conductivity. While this value is unrealistically large for magnetised
clusters, it serves our purpose here in highlighting the effects of
conduction, given also that our cluster is not particularly hot, so
that effects of conductivity can be expected to be weaker than in very
rich clusters.

In Figure~\ref{figprojections}, we show projections of the mass-weighted
temperature, X-ray emissivity, and gas mass density for all four
simulations. Each panel displays the gas contained in a box of side-length
$8.6\,{\rm Mpc}$ centred on the cluster. Comparing the simulations
with and without conduction, it is nicely seen how conduction tends to wipe
out small-scale temperature fluctuations. It is also seen that the outer
parts of the cluster become hotter when conduction is included.

These trends are also borne out quantitatively when studying radial
profiles of cluster properties in more detail. In
Figure~\ref{fig:cosmocompare}, we compare the cumulative baryonic mass
profile, the temperature profile, and the radial profile of X-ray emission
for all four simulations. Note that the innermost bins, for $R <
30\,{\rm kpc}$, may be affected by numerical resolution effects.
Interestingly, the temperature profiles of the
runs with conduction are close to being perfectly isothermal
in the inner parts of the cluster. While this does not represent a large
change for the adiabatic simulation, which is close to isothermal anyway,
the simulation with radiative cooling is changed significantly. Without
conduction, the radiative run actually shows a pronounced rise in the
temperature profile 
%towards the centre,
in the range of $100-200\,{\rm kpc}$,
as a result of compressional
heating when gas flows in to replace gas that is cooling out of the ICM in
a cooling flow. Only in the innermost regions, where cooling becomes
rapid, we see a distinct drop of the temperature.
Interestingly, conduction eliminates this feature in the
temperature profile, by transporting the corresponding heat energy
from the maximum both to
parts of the cluster further out and to the
innermost parts. The latter effect is probably small, however, 
because a smooth
decline in the temperature profile in the inner parts of the cluster, as
present in the ZN model, does not appear in the simulation.  As a
consequence, a strong conductive heat flow from outside to inside cannot
develop.

Conduction may also induce changes in the X-ray emission of the
clusters, which we show in the bottom panel of
Figure~\ref{fig:cosmocompare}.  Interestingly, the inclusion of
conduction in the adiabatic simulation has a negligible effect on the
X-ray luminosity. This is because in contrast to previous suggestions
\citep{Loeb02}, the cluster does not lose a significant fraction of
its thermal energy content to the outside intergalactic medium, and
the changes in the relevant part of the gas and temperature profile
are rather modest. We do note however that the redistribution of
thermal energy within the cluster leads to a substantial increase of
the temperature of the outer parts of the cluster.

For the simulations with cooling, the changes of the X-ray properties are
more significant. Interestingly, we find that allowing for thermal
conduction leads to a net {\em increase} of the bolometric luminosity of
our simulated cluster. The panel with the cumulative baryon mass
profile reveals that
conduction is also ineffective in significantly suppressing the condensation
of mass in the core regions of the cluster. In fact, it may even lead to
the opposite effect. We think this behaviour simply occurs because a
temperature profile with a smooth decline towards the centre, which would
allow the conductive heating of this part of the cluster, is not forming in
the simulation. Instead, the conductive heat flux is pointing primarily
from the inside to the outside, which may then be viewed as an additional
``cooling'' process for the inner cluster regions.

It is interesting to note that in spite of the structural effects that
thermal conduction has on the ICM of the cluster, it does
not affect its star formation history significantly.
In the two simulations that include cooling and star formation, the
stellar component of the mass profile 
(Figure~\ref{fig:cosmocompare}) does not show any sizeable difference
between the run including thermal conduction and the one without it.
This does not come as surprise as $90\%$ of the stellar content of
the cluster has formed before a redshift of $z=0.85$.
At these early times, the temperature of the gas in the protocluster
was much lower, such that conduction was unimportant.
In fact, for that reason, the stellar mass profiles
for both simulation runs coincide.

In summary, our initial results for this cluster suggest that
conduction can be important for the ICM, provided the
effective conductivity is a sizable fraction of the Spitzer
value. However, the interplay between radiative cooling and conduction
is clearly complex, and it is presently unclear whether temperature
profiles like those observed can arise in self-consistent
cosmological simulations. We caution that one should not infer too
much from the single object we examined here. A much larger set of
cluster simulations will be required to understand this topic better.

\section{Conclusions}
Hot plasmas like those found in clusters of galaxies are efficiently
conducting heat, unless electron thermal conduction is heavily
suppressed by magnetic fields. Provided the latter is not the case,
heat conduction should therefore be included in hydrodynamical
cosmological simulations, given, in particular, that conduction could
play a decisive role in moderating cooling flows in clusters of
galaxies. Such simulations are then an ideal tool to make reliable
predictions of the complex interplay between the nonlinear processes
of cooling and conduction during structure formation.

In this paper, we have presented a detailed numerical methodology for
the treatment of conduction in cosmological SPH simulations.  By
construction, our method manifestly conserves thermal energy, and we
have formulated it such that it is robust against the presence of
small-scale temperature noise.  We have implemented this method in a
modern parallel code, capable of carrying out large, high-resolution
cosmological simulations.

Using various test problems, we have demonstrated the accuracy and
robustness of our numerical scheme for conduction. We then applied our
code to a first set of cosmological cluster formation simulations,
comparing in particular simulations with and without conduction. While
these results are preliminary, they already hint that the
phenomenology of the coupled dynamics of radiative cooling and
conduction is complex, and may give rise to results that were perhaps
not anticipated by earlier analytic modelling of static cluster
configurations.

For example, we found that conduction does not necessarily reduce a
central cooling flow in our simulations; the required smoothly
declining temperature profile in the inner cluster regions does not
readily form in our cosmological simulations. Instead, the profiles we
find are either flat, or have a tendency to slightly rise towards the
centre, akin to what is seen in the cooling-only simulations. In this
situation, conduction may in fact lead to additional cooling in
certain situations, by either transporting thermal energy to the outer
parts, or by modifying the temperature and density structure in the
relevant parts of the cluster such that cooling is enhanced. This can
then manifest itself in an increase of the bolometric X-ray luminosity
at certain times, which is actually the case for our model cluster at
$z=0$.  Interestingly, we do not find that our cluster loses a
significant fraction of its thermal heat content by conducting it to
the external intergalactic medium.

In a companion paper \citep{Dolag04}, we analyse a larger
set of cosmological cluster simulations, computed with much higher
resolution and with more realistic sub-Spitzer conductivities. This
set of cluster allows us to investigate, e.g., conduction effects as a
function of cluster temperature and the influence of conduction on
cluster scaling relations.  While our first results of this work
suggest that conduction by itself may not resolve the cooling-flow
puzzle, it also shows that conduction has a very strong influence on
the thermodynamic state of rich clusters if the effective conductivity
is a small fraction of the Spitzer value or more. In future work, it
will hence be very interesting and important to understand the rich
phenomenology of conduction in clusters in more detail.

\section*{Acknowledgements}

The simulations of this paper were carried out at the Rechenzentrum
der Max-Planck-Gesellschaft, Garching.  K.~Dolag acknowledges support
by a Marie Curie Fellowship of the European Community program ``Human
Potential'' under contract number MCFI-2001-01227.

\bibliographystyle{mnras}
\bibliography{paper}

\end{document}